\documentclass[aps,prd,superscriptaddress,nofootinbib,twocolumn,notitlepage,longbibliography,10pt,floatfix]{revtex4-2}

\usepackage{float}
\usepackage{xcolor}
\usepackage{graphicx,epsf}
\usepackage{subfigure}
\usepackage{amsmath,amssymb,amsbsy,amstext,amsfonts,mathrsfs,latexsym,bm,physics,graphicx,xcolor}
\usepackage[colorlinks=true,urlcolor=blue,anchorcolor=blue,citecolor=blue,linkcolor=blue,linktocpage=true,pdfa=true]{hyperref}
\usepackage{orcidlink}

\begin{document}
\definecolor{orange}{rgb}{0.9,0.45,0}
%
%

%
\renewcommand{\t}{\times}
\long\def\symbolfootnote[#1]#2{\begingroup%
\def\thefootnote{\fnsymbol{footnote}}\footnote[#1]{#2}\endgroup}
\title{Mori-Zwanzig formalism for early cosmic inflation} 

\author{Ramin Hassannejad}
 \email{r.hassannejad@ut.ac.ir}
\affiliation{Department of Physics, University of Tehran, P.O. Box 14395-547 Tehran, Iran.}
\affiliation{Department of Physics, Shahid Beheshti University, 1983969411, Tehran, Iran.}
\begin{abstract} 
\noindent
The existence of fluctuations at the early stage of the universe provides enough confidence to rely on the averaging methods. However, the nonlinearity of general relativity makes it extremely difficult. There are different
methods to study inhomogeneous cosmology and provide a straightforward solution to the averaging problem
in cosmology, such as Buchert’s spatial averaging. I am attempting to study early cosmic inflation using Buchert equations and the Mori-Zwanzig projection operator formalism. Buchert equations and Mori-Zwanzig
formalism’s coarse-grained description are the geometrical source of early cosmic inflation through higherorder differential equations.
The theoretical results, while not an exact match, exhibit close agreement with the observational data, demonstrating the robustness of the model and its potential for further applications in cosmological analysis.

\end{abstract}


\maketitle
\vspace{0.8cm}

\section{Introduction} 
The concept of cosmic inflation, introduced by Alan Guth in 1981 \cite{Guth:1980zm, Guth:1981uk}, represents a paradigm shift in our understanding of the early universe. Guth proposed that a brief, exponential expansion occurred in the very early moments after the Big Bang, effectively addressing several long-standing cosmological issues, including the flatness problem, the horizon problem, and the absence of magnetic monopoles \cite{Guth:1979bh,Blau:1986cw,Guth:1985ya,Guth:1982pn, Preskill:1979zi}. His pioneering work laid the groundwork for what would become a cornerstone of modern cosmology \cite{weinberg2008cosmology}. 
The theoretical underpinnings of inflation were further elaborated in the seminal work of Linde \cite{Linde:1981mu}, who provided a comprehensive analysis of various inflationary potentials and their implications for the evolution of the universe \cite{Linde:1983gd, Linde:1986fd,Kachru:2003sx,Kallosh:2010xz,Kallosh:2010ug}. Linde's exploration of multiple inflationary models highlighted the vast landscape of possible inflationary scenarios \cite{Linde:1993cn,Linde:1997sj,Garcia-Bellido:1996mdl,Linde:1996gt, Linde:1991km}, paving the way for an expanding array of theoretical frameworks. This was complemented by the investigations of Guth, Linde, and others into the implications of inflation for cosmic microwave background (CMB) anisotropies and structure formation \cite{CMBPolStudyTeam:2008rgp,Garcia-Bellido:1996mdl,Greene:1997fu,Kofman:1994rk}. 
In subsequent years, the framework of inflation was significantly advanced by the contributions of Albrecht and Steinhardt \cite{Albrecht:1982wi, Albrecht:1982mp, Liddle:1994dx}, who introduced the notion of ``slow roll'' inflation. They demonstrated that inflation could be sustained through a scalar field, the inflaton, whose potential energy dominated its kinetic energy, leading to a prolonged phase of expansion. This model allowed for the generation of density perturbations through quantum fluctuations of the inflaton field, as detailed by Mukhanov and Chibisov \cite{Mukhanov:1981xt}, who showed how these fluctuations could evolve into classical perturbations that seeded the formation of large-scale structures in the universe \cite{Armendariz-Picon:1999hyi,Garriga:1999vw}. Moreover, one of the pioneering works in this field was by Starobinsky, who proposed an inflationary scenario driven by quantum corrections to general relativity. In his seminal paper, Starobinsky introduced a model based on an $R^{2}$
modification of Einstein’s equations, demonstrating that a de Sitter-like expansion could naturally emerge due to higher-order curvature terms \cite{Starobinsky:1980te}. This approach, often referred to as Starobinsky inflation, not only predicts a nearly scale-invariant spectrum of primordial perturbations but also remains in excellent agreement with modern observational data from the Cosmic Microwave Background (CMB) \cite{Planck:2018jri}. His work laid the foundation for later developments in inflationary cosmology, influencing a wide range of models and theoretical approaches \cite{Starobinsky:1986fx,Starobinsky:2007hu}.

The observational validation of inflationary cosmology has been bolstered by the findings of the WMAP \cite{WMAP:2003ivt} and Planck missions \cite{Planck:2018jri}, which have provided high-precision measurements of the CMB. These observations have confirmed the predictions of inflation regarding the scalar spectral index  and the amplitude of primordial fluctuations, reinforcing the theoretical framework developed in the foundational papers. The observed spectral index, , aligns remarkably well with the predictions of slow-roll inflationary models, lending significant credence to the inflationary paradigm. 
Despite its successes, the inflationary paradigm faces challenges within the standard model of cosmology. These include the initial conditions problem \cite{Goldwirth:1991rj,Kaloper:2002cs, Linde:2017pwt,Linde:1985ub}, which questions the likelihood of specific field configurations necessary for inflation, and the measure problem, which complicates probability assignments to different inflationary outcomes. The transition from quantum fluctuations to classical density perturbations remains an active area of research \cite{Maldacena:2002vr,Starobinsky:1986fx, Starobinsky:1994bd}. These unresolved issues motivate the exploration of alternative approaches and novel theoretical tools. Moreover, studying the averaging problem in cosmology has the potential to solve some problems in cosmology, which have been considered in different research projects \cite{Buchert:2001sa, Buchert:1999er, Buchert:2019mvq, Vrugt:2021sfu,Wiltshire:2007fg,Zalaletdinov:2008ts,Ginat:2020nqw,Garfinkle:2023vzf}.

The study of memory-dependent equations of state (MDES) in cosmology presents a novel approach to understanding the evolution of the universe. Traditional equations of state typically assume an instantaneous response between pressure and energy density, but recent advances in nonequilibrium thermodynamics suggest that the behavior of cosmic fluids may depend on their historical states \cite{Vrugt:2021sfu}. This memory effect, rooted in integral formulations, introduces corrections that could have significant implications for early-universe dynamics and inflationary models. 
The key motivations for incorporating memory effects in cosmology stems from the Mori-Zwanzig formalism \cite{Mori:1965oqj,Zwanzig:1960gvu}, a powerful framework in statistical mechanics that effectively separates relevant and irrelevant degrees of freedom \cite{te2020projection}. By applying this formalism to the Friedmann equations, we obtain modified dynamical equations that naturally accommodate memory terms. These modifications may address unresolved issues in cosmological models, such as the nature of inflation, the behavior of dark energy, and deviations from the classical perfect-fluid approximation.  This mathematical framework enables the derivation of effective equations of motion for relevant observables by systematically accounting for neglected degrees of freedom \cite{Vrugt:2021sfu}. By focusing on the most pertinent variables and integrating out the less relevant ones, the formalism can simplify the complex dynamics of inflation. Its application to cosmic inflation bridges microscopic dynamics and macroscopic behavior, potentially offering insights into the quantum-to-classical transition of primordial perturbations and the origin of stochastic effects during inflation.

In this work, we systematically develop a memory-dependent equation of state and explore its consequences for cosmic evolution. We demonstrate that incorporating memory effects can give rise to an inflationary phase that is in close agreement with observational constraints from Planck, BAO, BK18, and the Atacama Cosmology Telescope (ACT), thereby reinforcing the compatibility of our theoretical framework with current cosmological observations.
 Furthermore, we analyze the impact of these effects on slow-roll parameters and derive new conditions under which inflation occurs. Our results suggest that the inclusion of memory terms in the equation of state provides a richer and more nuanced description of the early universe, potentially offering insights into deviations from standard cosmological predictions.
\\
The structure of this paper is as follows: In Section II, we review the theoretical foundations of stress, strain, and viscoelasticity in the context of memory-dependent equations. Section III presents the derivation of the memory-dependent equation of state and its incorporation into the Friedmann equations. In Section IV, we apply the Mori-Zwanzig formalism to derive a generalized evolution equation for the Hubble parameter. Section V explores the implications of our model for inflationary dynamics and compares its predictions with observational data. Section VI introduces the memory-dependent scalar field, its associated potential function, and innovative approaches for developing memory-dependent inflation models. Finally, we summarize our key findings and discuss potential future directions in Section VII.
\section{Memory-Dependent Stress and Strain}
In this section, we discuss several well-known physical models that exhibit memory effects in their evolution, such as viscoelasticity.\\ Viscoelasticity, the interaction between elastic and viscous behaviors, represents a significant area of study, not only due to its relevance in practical applications and technological advancements but also from a theoretical perspective. All natural materials display a combination of viscous and elastic properties when subjected to deformation \cite{Lakes1998, Christensen2003,GutierrezLemini2013, Andrade:2019zey}.
\subsection{Stress and Strain}
Let’s first take a look at the basic and classic definitions of stress and strain in Newtonian mechanics:
Stress and strain are  concepts in mechanics that describe the response of materials to external forces \cite{young2002roark, openstax2020university}.\\ \textit{Stress} ($\sigma$) is a measure of the internal force per unit area within a material, defined mathematically as:
\begin{equation}
	\sigma = \frac{F}{A},
\end{equation}
where $F$ is the applied force, and $A$ is the cross-sectional area over which the force acts. Depending on the direction and nature of the applied load, stress can be categorized as normal stress (acting perpendicular to the surface) or shear stress (acting tangentially to the surface).\\\\
\textit{Strain} ($\varepsilon$) quantifies the deformation of a material relative to its original dimensions \cite{young2002roark, openstax2020university}. Normal strain, associated with stretching or compressing, is expressed as:
\begin{equation}
	\varepsilon = \frac{\Delta L}{L_0},
\end{equation}
where $\Delta L$ is the change in length and $L_0$ is the original length. 
Under small deformations within the elastic regime, the relationship between stress and strain is governed by \textit{Hooke’s law} \cite{jastrzebski1959nature}. For normal stresses and strains, this relationship is expressed as:
\begin{equation}
	\sigma = E \varepsilon,
\end{equation}
where $E$ is the Young’s modulus, a material property that characterizes stiffness. 
Understanding stress and strain is crucial for analyzing material behavior under various loading conditions, providing the foundation for more complex models in solid mechanics.
\subsection{Linear Viscoelasticity}
Linear viscoelasticity describes materials whose stress-strain relationship depends on time (viscoelastic behavior) but remains linear under small deformations \cite{tschoegl1977boltzmann,Coleman1964ThermodynamicsOM,boltzmann1874nachwirkung, Boltzmann_2012}. It is a fundamental framework used in mechanics to model materials such as polymers, biological tissues, and metals at certain temperatures. 
The behavior of linear viscoelastic materials \cite{coleman1961foundations} can be described using integral equations. These equations express stress or strain as a function of time, accounting for the material's memory.\\
The integral function for strain, as described in \cite{Irgens2008}, is expressed as:
\begin{equation}
	\label{dkfjbwbfw}
	\boldsymbol{\epsilon}(t) = \boldsymbol{\sigma}(0) K(t)
	+ \int_{0}^{t} K(t - \tau) \frac{d\boldsymbol{\sigma}(\tau)}{d\tau} \, d\tau
\end{equation}
where $K(t-\tau)$ is the memory function, describing how past stress  influence the current strains. 
Moreover, the stress integral, as outlined in \cite{Irgens2008}, can be written as:
\begin{equation}\label{dkjfejf}
	\boldsymbol{\sigma}(t)=\boldsymbol{\epsilon}(0)E(t)+\int_{0}^{t} E(t-\tau)\frac{d \boldsymbol{\epsilon}(\tau)}{d\tau}d\tau	
\end{equation}
where $E(t-\tau)$ is the memory function, describing how past strains influence the current stress. 
The above integral equations indicate that the current values of the functions on the left-hand side depend on the integral functions on the right-hand side, which sum up the history of the evolution of the given system, such as a solid or fluid. 
The various physical properties and applications of these equations have been extensively studied over the past decades; for some examples, see \cite{coleman1961foundations, kerr1964elastic, Andrade:2019zey}.
\subsection{Other Models of Viscoelasticity}
\subsubsection{Bernstein–Kearsley–Zapas model}

The Bernstein–Kearsley–Zapas (BKZ) model \cite{bernstein1963stress} is a  model used in rheology to describe the nonlinear viscoelastic behavior of polymer melts and other complex fluids. This model is based on the theory developed by Green and Rivlin \cite{Green1957,Green1959part2,Green1959}, which was formulated to study the mechanics of nonlinear materials with memory.  \\
In Appendix \eqref{dekwdwkj}, I provided a detailed explanation of the main concepts and calculations related to the BKZ model, ultimately arriving at Eq.~\eqref{cdnwdlfwd}. However, due to the complexity of this equation, we can simplify it by considering the special case where $n = 1$ (see \cite{bernstein1963study}). In this scenario, Eq.~\eqref{cdnwdlfwd} reduces to
\begin{align}\label{dcwdjkcw}
 p \delta^{ij}=& -\sigma^{ij}(t)+ F^i_{\mu}(t) F^j_\nu(t) \big(\chi^{\mu\nu} +\chi^{\mu\nu\alpha\beta}E_{\alpha\beta}(t) \big)
\notag\\
&-F^i_{\mu}(t) F^j_\nu(t) \int_{-\infty}^{t}K^{\alpha\beta\mu\nu}(t-s)E_{\alpha\beta}(s)ds
\end{align}
This equation indicates that the stress tensor at time $t$ depends on the entire history of the system over the interval $-\infty < s < t$. In fact, the stress tensor in the BKZ model includes a memory term. 
\\\\
Furthermore, Eq.~\eqref{dcwdjkcw} can be expressed in an alternative form, as shown in \cite{bernstein1964thermodynamics} in the context of perfect elastic fluids. I provided a detailed explanation of the concept and calculations for this scenario in Appendix~\ref{dkldefkddfni}, ultimately arriving at Eq.~\eqref{dfgnbgfbfg}, which can be written as follows
\begin{align}\label{rfgfrgfrg}
	p \delta^{ij}= -\sigma^{ij}(t) 
	+\rho T \int_{-\infty}^{t}&ds~F^{i}_{\alpha}(t,s)F^{j}_{\beta}(t,s)b_{T}(s)
	\notag\\
	&\times S^{\alpha\beta}\Big[\boldsymbol{E}(t,s),\int^{t}_{s}b_{T}(\eta)d\eta\Big]
\end{align}
where $\rho$ represents the energy density, $T$ denotes the temperature, and the quantity $b_{T}$ depends on absolute temperature. This illustrates that, similar to the well-known equations of state in the thermodynamics of equilibrium systems, the pressure is a function of temperature and energy density. However, Eq.~\eqref{rfgfrgfrg} includes a memory term, and it could be useful to consider an equation of state where the pressure depends on energy density and temperature within this framework. 
\subsubsection{Wagner model}	
The Wagner model is often viewed as a more straightforward application of the Bernstein–Kearsley–Zapas model \cite{bernstein1963stress, bernstein1963study, bernstein1964thermodynamics, kaye1962nonnewtonian}.  It was introduced by Wagner in several papers \cite{Wagner1976, Wagner1977, Wagner197922, Wagner1980, Wagner1992, Wagner1998}. The stress tensor, expressed as a function of a memory term, is given by
\begin{equation}\label{wdkwkjdwjd}
\boldsymbol{\sigma}(t)=-p\bold{I}+\int_{-\infty}^{t}M(t,\tau, \boldsymbol{A}) \bold{A}(\tau)d\tau
\end{equation}
Here, $\boldsymbol{\sigma}(t)$ represents the stress tensor at time $t$, $\bold{I}$ is the identity tensor, $p$ denotes the isotropic pressure contribution, $\bold{A}(\tau)$ is the Finger tensor at time $\tau$ (prior to time $t$), and $M(t,\tau, \boldsymbol{A})$ is the memory function.\\ 
In the rubberlike liquid equation, the memory function depends solely on the time interval $(t - \tau)$ \cite{Winter1978} and is given by
\begin{equation}
M(t-\tau)=\sum_{i=1}^{N}\frac{\theta_{i}}{\lambda_{i}}\exp\bigg(-\frac{t-\tau}{\lambda_{i}}\bigg)
\end{equation}
where $\theta_{i}$ represents the moduli of linear viscoelasticity, and $\lambda_{i}$ denotes the time constants of linear viscoelasticity (relaxation time).
\subsubsection{Pressure as a Function of Oblivion and Memory Terms}
We know that
both stress and pressure share the same units, but their physical interpretations differ. Stress relates to internal forces within a solid, whereas pressure pertains to external forces or forces within fluids. In Eqs.~\eqref{dkjfejf} and \eqref{wdkwkjdwjd}, we demonstrated that stress can be expressed as a memory-dependent function. However, one may wonder what would happen if a similar approach were applied to formulate a pressure equation with memory dependence.  For instance, the integral equation given in Eq.~\eqref{wdkwkjdwjd} can be expressed as
\begin{align}\label{rgrggrtrgh}
	p\bold{I}=\underbrace{-\boldsymbol{\sigma}(t)}_{\text{oblivion term}}
	+\underbrace{\int_{-\infty}^{t}M(t,\tau, \boldsymbol{A}) \bold{A}(\tau)d\tau}_{\text{memory term}}
\end{align}
I divided the right-hand side into two parts: the first part, which is solely a function of the current time \(t\), is referred to as the \textit{oblivion} term, while the second part, which depends on both the present time and the historical state of the matter with respect to time \(\tau\), is called the \textit{memory} term. \\
To gain insight into this definition, let us expand the pressure equation \eqref{dcwddcwdidcw} around $\Sigma=0$, which yields the following expression
\begin{equation}\label{wdfjwdfw}
p\simeq p_{eq}+\frac{\partial p_{eq}}{\partial h}\int^{t}_{-\infty}S\big[\boldsymbol{E}(t,\tau),B(t, \tau)\big]b_{T}(\tau)d\tau	
\end{equation}
This equation shows that there is a memory term in the pressure function close to the equilibrium state. The oblivion term represents the pressure at equilibrium, while the memory term includes the pressure derivative and an integral term.
\\\\
In this paper, I study this type of pressure equation to illustrate how it can enhance our understanding of the early universe, particularly during the period of cosmic inflation. I define an equation of state that incorporates both oblivion and memory terms and investigate a more realistic example of such an equation using the Mori–Zwanzig formalism. Consequently, we explore cosmological scenarios where a memory-dependent equation of state may provide valuable insights.
\section{Memory-Dependent Equation of State}

The calculations and definitions related to stress and strain discussed in the previous sections are based on nonequilibrium thermodynamics, which is a broad area of study in physics and engineering. These topics are discussed to illustrate that memory effects, which account for the past history of a system, can naturally appear in equations involving pressure as well as in general thermodynamic equations of state.

In this study, we consider the Einstein gravitational field equations with an effective  fluid energy-momentum tensor as follows
\begin{equation}\label{wdwdwed}
	G_{\mu\nu}=T^{\text{eff}}_{\mu\nu}, \quad \text{with } \quad T^{\text{eff}}_{\mu\nu}=(\rho+p^{\text{eff}})u_{\mu}u_{\nu}+p^{\text{eff}}g_{\mu\nu}
\end{equation}
where we assume $8\pi G_{0}=1$. 
The source of this effective fluid may originate either from geometrical contributions on the left-hand side of Einstein’s field equations (see Ref.~\cite{Khodabakhshi:2024med}), or from additional source terms encoded in the stress–energy tensor on the right-hand side.\\

Indeed, both the geometrical terms on the left-hand side and the matter terms on the right-hand side can be defined as a kind of effective energy-momentum term. To take an exact look at this, we want to consider Jacobson's theory about the origin of Einstein's gravitational field equations. In 1995, Ted Jacobson showed that the Einstein field equations can be derived from thermodynamic principles \cite{Jacobson:1995ab}. He demonstrated that by assuming the proportionality of entropy to horizon area and applying the fundamental relation $\delta Q = T\,dS$
for all local Rindler horizons, the Einstein equations emerge naturally as an equation of state for spacetime. \\

Building on this, Eling, Guedens, and Jacobson (see Ref.~\cite{Eling:2006aw}) extended the derivation to include curvature corrections to the entropy that depend on the Ricci scalar. They showed that such corrections require a \emph{non-equilibrium} thermodynamic treatment, where the entropy balance relation takes the form
\begin{equation}\label{dcdcwdc}
	dS = \frac{\delta Q}{T} + d_i S,
\end{equation}
with $d_i S$ representing internal entropy production, analogous to a bulk viscosity term. By enforcing local energy-momentum conservation, they derived the corresponding modified field equations, showing that even in the presence of higher curvature contributions, the thermodynamic equation of state remains consistent with the generalized gravitational dynamics. They showed that in the presence of internal entropy production $d_i S$, arising from non-equilibrium effects such as bulk viscosity associated with the horizon, the gravitational field equations generalize to the case where the Lagrangian is a function of the Ricci scalar, $f(R)$. \\
Thus, the existence of the viscosity term $d_i S$ in Eq.~\eqref{dcdcwdc} leads to a modification of gravity with a Lagrangian corresponding to $f(R)$ theory. This modification affects the left-hand side of Einstein's field equations (see Ref.~\cite{Khodabakhshi:2024med}), where the additional terms can be interpreted as an effective energy-momentum tensor with an effective pressure contribution, according to Eq.~\eqref{wdwdwed}. Moreover, the presence of an effective pressure induced by the viscosity term is demonstrated in Appendix~\ref{appendixcc}, where it is shown that the viscosity term contributes to the effective pressure as  
\begin{equation}\label{jsdjwdjs}
	p^{\text{eff}} = p_{e} + \Pi .
\end{equation}
This effective pressure equation is similar to Eq.~\eqref{wdfjwdfw}, as both originate from viscosity and non-equilibrium thermodynamics. Hence, it is expected that the pressure term $\Pi$ is given by an integral form and exhibits a memory effect. It can be defined as a general equation of state that incorporates an integral term with a memory effect. A Memory-Dependent Equation of State (MDES) is a formulation in which the current state of a system, such as pressure, energy, temperature, or other state variable, is influenced not only by its present conditions but also by its past states or history. The equation often involves integral terms, \textit{convolution operators} \footnote{Given two functions $f, g: \mathbb{R}^n \to \mathbb{R}$, their convolution is defined as 
		\[
		(f * g)(x) = \int_{\mathbb{R}^n} f(y) g(x - y) \, dy,
		\]
provided that the integral is well-defined. Convolution is a fundamental operation in analysis, particularly in the study of partial differential equations (PDEs), harmonic analysis, and signal processing. It describes how a function is "smeared" by another, often used to smooth data, solve differential equations, and analyze linear systems.}, or time-delayed functions to model the influence of past states.\\ Therefore, by taking into account Eqs.~\eqref{wdkwkjdwjd}, \eqref{rgrggrtrgh}, and \eqref{wdfjwdfw}, the memory-dependent equation of state can be written as
\begin{align}\label{djcjcwdowo}
	p^{\text{eff}}(t) =\underbrace{f\big(\rho(t), T(t)\big)}_{\text{oblivion term}} + \underbrace{\int_{0}^{t} g\big(\rho(t-s), T(t-s)\big) \, ds}_{\text{memory term}}
\end{align}	
where $g\big(\rho(t-s), T(t-s)\big)$ represents the memory effect. The pressure at time $t$ is given by the integral of all contributions from earlier times $s < t$.\\
This memory-dependent equation of state has significant capabilities, which I will explore in the early universe, though it can also be studied in other periods of cosmology and gravitational physics.
\\
To study the MDES in cosmology, one needs to solve Einstein’s field equations. Accordingly, the corresponding Friedmann equations are given as
\begin{align}\label{dcdlkwddlwnd}
&H^2(t)=\frac{\rho(t)}{3}\\
&\dot{H}(t)=-\frac{\rho(t)+p^{\text{eff}}(t)}{2}
\label{dwdwdcd}
\end{align}
where the Hubble parameter is $H(t)=\dot{a}(t)/a(t)$, and $\dot{H}(t)=dH(t)/dt$.\\
The continuity equation for the FLRW space-time, in the presence of the effective energy-momentum tensor given in Eq.~\eqref{wdwdwed}, can be written as
\begin{equation}\label{contin}
	\dot{\rho} + 3H\left(\rho + p_{e}+\Pi\right) = 0.
\end{equation}
Substituting the equation of state from Eq.~\eqref{djcjcwdowo} into the Friedmann equations yields a field equation with significant predictions. I will demonstrate this in the following sections.

\subsection{Pressure Including Memory and Oblivion Terms}

Equation~\eqref{djcjcwdowo} represents the most general form of the memory-dependent equation of state (MDES), in which the pressure is allowed to depend on both the energy density and the temperature. Such a formulation is broad enough to be applied across different areas of physics and cosmology. However, in cosmology, the conventional equations of state are typically expressed only as functions of the energy density. To remain consistent with this framework and for the sake of simplicity, we restrict our analysis to the case where the MDES depends solely on the energy density, while still incorporating a non-local memory contribution. In this simplified scenario, the pressure is written as
\begin{equation}\label{efvfrvfdv}
	p^{\text{eff}}(t) = f\big(\rho(t)\big) + \int_{0}^{t} g\big(\rho(t-s)\big), ds ,
\end{equation}
where the first term, $f(\rho)$, is interpreted as the oblivion (instantaneous) contribution, while the second integral term introduces the memory effect through the kernel $g$.
\\
The first term, $f(\rho)$, can take various well-established functional forms, some of which play a crucial role in the study of gravitational collapse in massive objects (see Refs.~\cite{Hassannejad:2024cbu, Hassannejad:2023lrp, Shojai:2022pdq}).
\\ 
Substituting  pressure Eq.~\eqref{efvfrvfdv} into Eq.~\eqref{dcdlkwddlwnd} leads to the following equation
\begin{equation}
	\ddot{H}=-\frac{1+f^{\prime}\big(\rho\big)}{2}\dot{\rho}-\frac{1}{2}\frac{d}{dt}\int_{0}^{t} g\big(\rho(t-s)\big) \, ds.
\end{equation}
By replacing the equation $ g\big(\rho(t-s)\big) = \beta e^{-\zeta s} F\big(\rho(t-s)\big)$ along with the continuity equation  
$
\dot{\rho} = -3H(\rho + p^{\text{eff}})
$  
into the above second-order differential equation, we obtain
\begin{align}\label{dcncwdkchw}
	\ddot{H}&+3\dot{H}H\big(1+f^{\prime}(\rho)\big)+\zeta\Big(\dot{H}+\frac{3}{2}H^2\Big)
	\notag\\
	&+\frac{\zeta f(\rho)+\beta F(\rho)}{2}=0
\end{align}
where $\beta$ and $\zeta$ are two constants whose values can be determined through a precise analysis.
\\
This second-order differential equation can be solved exactly to determine the Hubble parameter. However, we are now primarily interested in studying early cosmic inflation using this approach. By applying the slow-roll conditions from Eqs.~\eqref{dfkwddfbwqe} and \eqref{gbfgfdc}, Eq.~\eqref{dcncwdkchw} simplifies to
\begin{equation}
	\dot{H} H\big(1+f^{\prime}(\rho)\big)+\frac{\zeta }{2}H^2+\frac{\zeta f(\rho)+\beta F(\rho)}{6}\simeq 0.
\end{equation}	
This equation can be written in the form of the Friedmann equations
\begin{equation}\label{dfbrefgsg}
	\dot{H}=-\frac{1+\omega(\rho)}{2}\rho,\quad \text{with}\quad \omega(\rho)=\frac{p^{\text{eff}}}{\rho}
\end{equation}
where $\omega(\rho)$ is expressed as
\begin{align}\label{rgthetrhr}
	\omega(\rho) \simeq-1+\frac{\zeta }{\sqrt{3\rho}\big(1+f^{\prime}(\rho)\big)}+\frac{\zeta f(\rho)+\beta F(\rho)}{\sqrt{3}\rho^{3/2}\big(1+f^{\prime}(\rho)\big)}.
\end{align}
The equation of state parameter approaches  $\omega \simeq-1$, under the following conditions
\begin{align}\label{dkjcdjk}
	&f^{\prime}(\rho)\gg \frac{\zeta}{\sqrt{3\rho}}-1,
	\\
	&f^{\prime}(\rho)\gg \frac{\zeta f(\rho)+\beta F(\rho)}{\sqrt{3}\rho^{3/2}}-1.
	\label{dllnlwedn}
\end{align}
For appropriately chosen forms of $f(\rho)$ and $F(\rho)$, these conditions can be examined in the asymptotic regimes of both high and low energy density.  
For instance, consider the case where $f(\rho) = \rho$ and $F(\rho) = \sqrt{\rho}$.  
Substituting these forms into Eq.~\eqref{rgthetrhr} yields
\begin{equation}\label{dwfnedf}
	\omega(\rho) \simeq -1+\frac{\zeta}{\sqrt{3\rho}}+\frac{\beta}{2\sqrt{3}\rho}
\end{equation}
for this equation of state, the conditions in Eqs.~\eqref{dkjcdjk} and \eqref{dllnlwedn} can be written as
\begin{align} 
	\rho \gg \frac{\zeta^2}{3}, \quad \text{and} \quad \rho \gg \frac{\beta}{2\sqrt{3}}.
\end{align}
In the high-energy-density regime, the equation of state given by Eq.~\eqref{dwfnedf} approaches $\omega \simeq -1$, resulting in an exponential growth of the scale factor,
$
a(t) \propto e^{\eta t}.
$
This behavior indicates the occurrence of an inflationary phase at high energy densities, similar to that of the early universe.
 \\
Let's review what happened above: I introduced a memory-dependent equation of state and substituted it into the Friedmann equations, which led to a second-order differential equation. Applying the slow-roll conditions reduced this to a first-order differential equation. Finally, rewriting it in the form of the Friedmann equations resulted in an equation of state that predicts an inflationary phase in the high-energy-density limit. In the following sections, as I delve into the Mori-Zwanzig approach, the above method will become more apparent.
\subsection{Pressure Including Only the Memory Term}
Another possible scenario is to set the oblivion term to zero, allowing the components of the memory-dependent equation of state (MDES) to be written as
\begin{align}\label{dnkcwjdfow}
	&f\big(\rho(t), T(t)\big)=0, \notag\\
	&g\big(\rho(t-s), T(t-s)\big)=\frac{2\kappa}{\sqrt{3}}\, e^{-\zeta s}F\big(\rho(t-s)\big),
\end{align}
which enables us to study the cosmological dynamics driven solely by the memory term. 
Substituting these equations into Eq.~\eqref{djcjcwdowo} results in
\begin{equation}\label{frgfrgfrg}
	p^{\text{eff}}(t) =  \frac{2\kappa}{\sqrt{3}}\int_{0}^{t} ds e^{-\zeta s}F\big(\rho(t-s)\big) .
\end{equation}
this pressure is influenced by the system's history over time. Substituting it into the Friedmann equations \eqref{dcdlkwddlwnd} and \eqref{dwdwdcd} yields
\begin{align}\label{rgergrg}
	\ddot{H}(t)=&-\frac{\dot{\rho}}{2}-\frac{\kappa}{\sqrt{3}}\Big(e^{-\zeta t}F\big(\rho(0)\big)\Big)\notag\\
	&+\frac{\kappa}{\sqrt{3}}\int_{0}^{t} ds e^{-\zeta s}\frac{\partial F\big(\rho(t-s)\big)}{\partial s}
\end{align}
after some calculations, the above equation simplifies to
\begin{align}\label{feferfe}
	\ddot{H}+3\dot{H}H+\zeta\Big(\dot{H}+\frac{3}{2}H^2\Big)+\frac{\kappa}{\sqrt{3}}F\big(\rho(t)\big)=0.
\end{align}

Equation~\eqref{feferfe} encapsulates the effective dynamics of the Hubble parameter when the pressure is governed solely by a memory-dependent contribution. The terms 
$\ddot{H}+3H\dot{H}$ reflect the standard dynamical structure familiar from the Friedmann equations, while the additional terms proportional to $\zeta$ arise directly from the exponential kernel in the memory function, effectively introducing a damping mechanism in the evolution of $H(t)$. The final contribution, proportional to $\kappa F(\rho)$, encodes the explicit functional dependence of the memory effect on the energy density. Taken together, these modifications illustrate how incorporating memory effects alters the standard cosmological dynamics while preserving a closed and local form for the evolution of the Hubble parameter.
\\

Applying the slow-roll conditions from Eqs.~\eqref{dfkwddfbwqe} and \eqref{gbfgfdc} to Eq.~\eqref{feferfe} leads to

\begin{align}\label{fdcedfqwefc}
	\dot{H}+\frac{\zeta}{2}H+\frac{\kappa}{3\sqrt{3}H}F\big(\rho(t)\big)\simeq 0
\end{align}
it is straightforward to express the above equation in the form of the Friedmann equations
\begin{align}\label{efergergre}
	\dot{H}=-\frac{1+\omega(\rho)}{2}\rho, \quad \text{with}\quad \omega(\rho)\simeq-1+\frac{\zeta}{\sqrt{3\rho}}+\frac{2\kappa}{3}\frac{F(\rho)}{\rho^{3/2}}.
\end{align}
For all functions of $F(\rho)$ where $\lim_{\rho \to \infty} F(\rho)/\rho^{3/2} \to 0$, the equation \eqref{efergergre} tends to equation of state with $\omega \simeq -1$, and it can be a possible model to describe the early cosmic inflation. For example, if we take $F(\rho) = \rho^n$, the given condition is satisfied for $n < 3/2$.

\subsection{Energy Conditions}
\begin{figure}[h] 
	\centering 
	\includegraphics[width=3.7in]{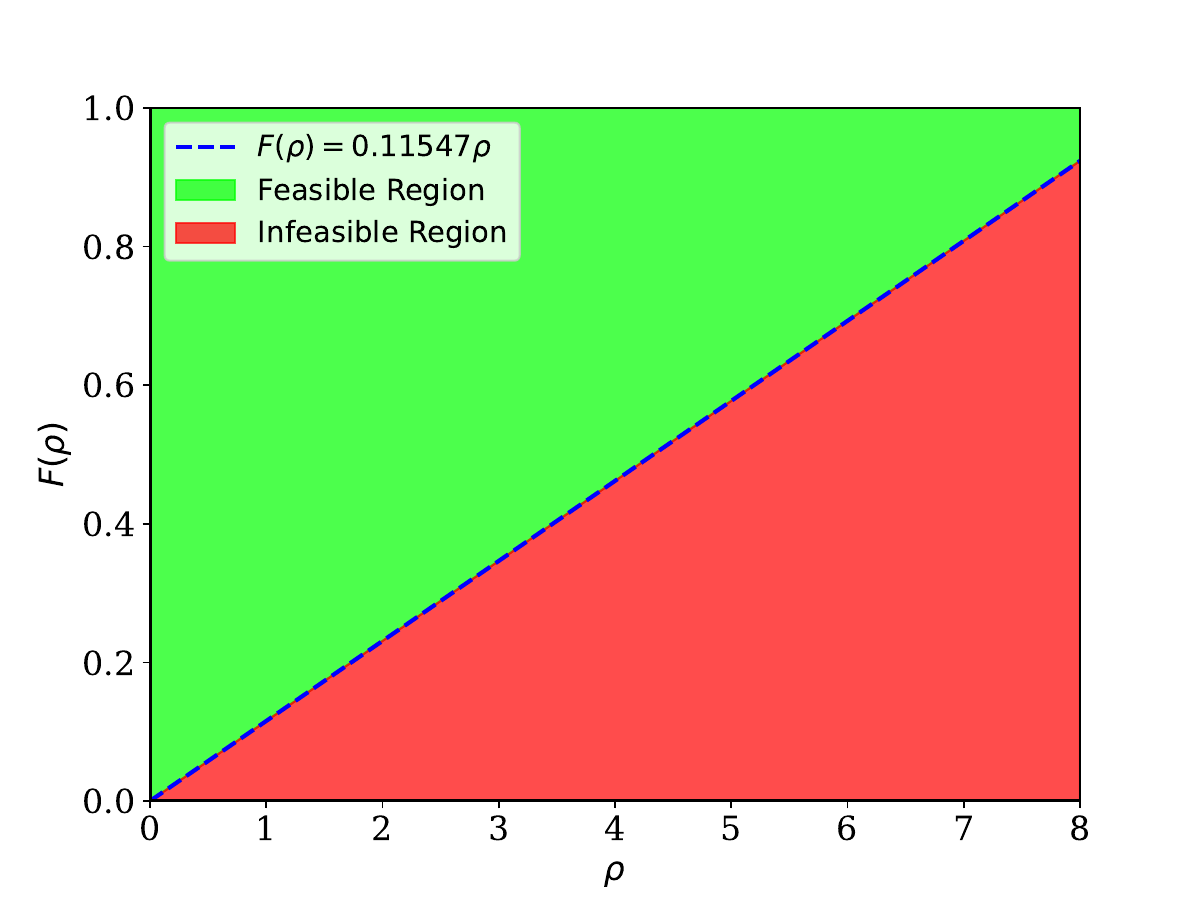}
	\caption{\label{fighhrhop}	
This figure illustrates the regions where the NEC is satisfied and violated. It is based on the condition in Eq.~\eqref{wdckwkefoei} for the values $\zeta=-0.04$ and $\kappa=0.3$. The NEC is satisfied in the green area and violated in the red area.}
\end{figure}

Another constraint on the possible forms of $F(\rho)$ arises from the energy conditions in general relativity, which are defined as follows
\begin{itemize}
	\item \textbf{NEC: Null Energy Condition.}  
	The NEC stipulates that $ T^{\text{eff}}_{\mu\nu} k^\mu k^\nu \geq 0$ for any null vector $k_\mu$ satisfying $k^\mu k_\mu = 0$. For a perfect fluid, this condition simplifies to
	\begin{equation}
	\rho + p^{\text{eff}} \geq 0
	\end{equation}
for the equation of state in Eq.~\eqref{efergergre}, the NEC takes the form
\begin{equation}
	\label{wdckwkefoei}
F(\rho)\geq \frac{\sqrt{3}|\zeta|}{2\kappa}\rho
\end{equation}
I assume that $\kappa > 0$ and $\zeta < 0$, as these will be useful in the following sections. Fig. \ref{fighhrhop} shows the condition described in Eq.~\eqref{wdckwkefoei}. The NEC is satisfied in the green area and not satisfied in the red area.
\end{itemize}

\begin{itemize}
	\item \textbf{WEC: Weak Energy Condition.}  
	The WEC requires that $ T^{\text{eff}}_{\mu\nu} t^\mu t^\nu \geq 0$ for any timelike vector $t_\mu$ satisfying $t^\mu t_\mu < 0$ (often normalized as $t^\mu t_\mu = -1$). By continuity, this condition extends to null vectors $k_\mu$. For a perfect fluid, the WEC reduces to the following
	\begin{equation}
	\rho \geq 0 \quad \text{and} \quad \rho + p^{\text{eff}} \geq 0.
	\end{equation}
The energy density is positive, indicating that the first condition is satisfied. The second condition is identical to the NEC condition. Based on the equation of state in Eq.~\eqref{efergergre}, the WEC can be expressed as
	\begin{equation}
		\label{fvfev}
		F(\rho)\geq \frac{\sqrt{3}|\zeta|}{2\kappa}\rho
	\end{equation}
The behavior of the condition described above is illustrated in Fig. \ref{dfsdfv}.
\end{itemize}

\begin{itemize}
	\item \textbf{DEC: Dominant Energy Condition.}  
	The DEC ensures that the energy-momentum current density $P_\mu = T^{\text{eff}}_{\mu\nu} t^\nu$, as measured by an observer with a timelike and future-directed 4-velocity $t^\nu$, remains causal. This requires $P^\mu$ to be a future-directed, non-spacelike vector, which imposes the conditions  $P^\mu t_\mu \leq 0$ and $P^\mu P_\mu \leq 0$. For a perfect fluid, this translates into the following constraint 
	\begin{equation}\label{dmdjdj}
		\rho \geq |p^{\text{eff}}|\,.
	\end{equation}
	\begin{figure}[h] 
		\centering 
		\includegraphics[width=3.6in]{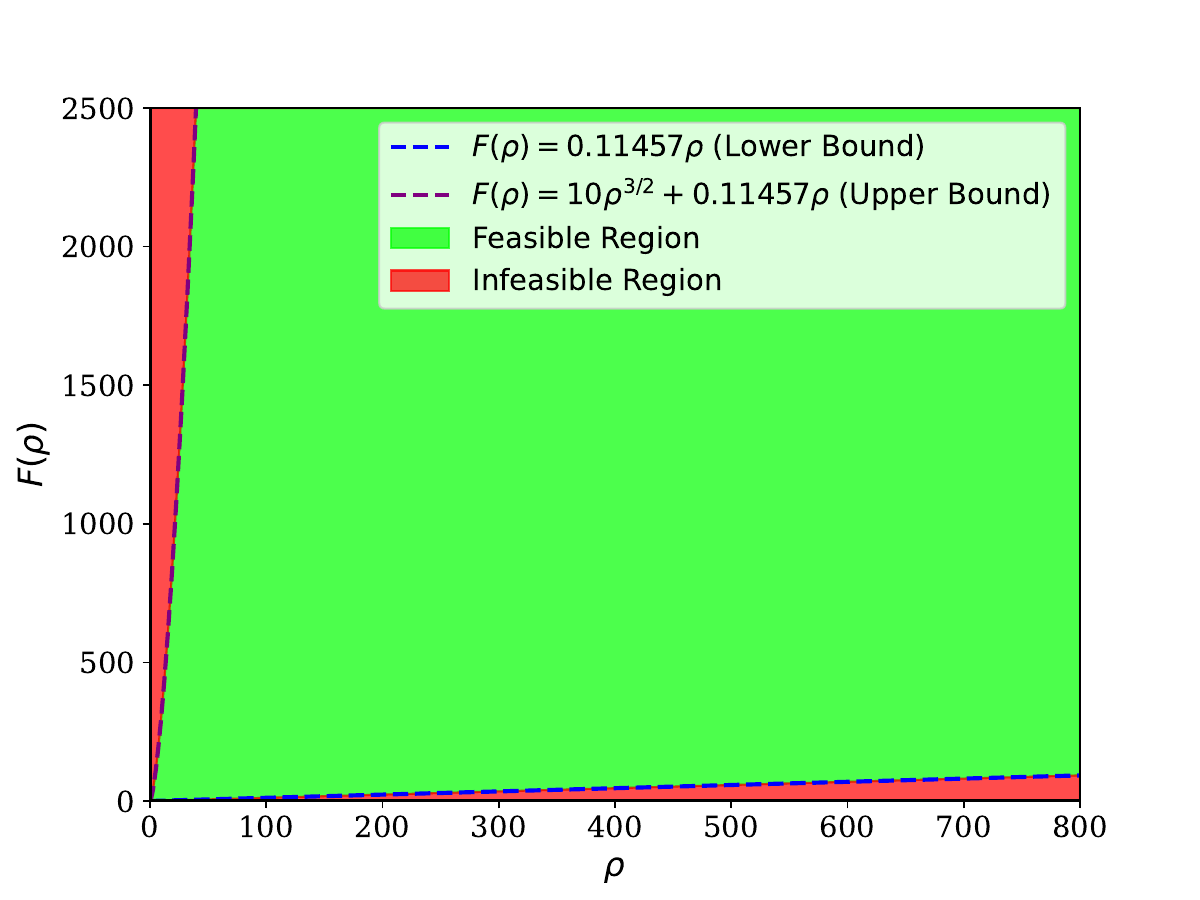}
		\caption{\label{dfsdfv}	
	This figure displays the regions where the WEC is satisfied and violated.  It is generated according to the condition in Eq.~\eqref{wdnddcwdwe} for the values $\zeta=-0.04$ and $\kappa=0.3$. The WEC is satisfied in the green area and violated in the red area.
		}
	\end{figure}
Under this condition, applying the equation of state, Eq.~\eqref{efergergre}, gives
\begin{equation}
	\bigg| -1+\frac{\zeta}{\sqrt{3\rho}}+\frac{2\kappa}{3}\frac{F(\rho)}{\rho^{3/2}}\bigg|\leq 1
\end{equation}
through detailed calculations, we arrive at 
	\begin{equation}\label{wdnddcwdwe}
		\frac{\sqrt{3}|\zeta|}{2\kappa}\rho \leq F(\rho)\leq \frac{3}{\kappa}
		\rho^{3/2}+\frac{\sqrt{3}|\zeta|}{2\kappa}\rho.
	\end{equation}
Fig.~\ref{dfsdfv}  ~provides a visual representation of the condition described above. The DEC is satisfied in the green region, while it is violated in the red regions.	
	\item \textbf{SEC: Strong Energy Condition.}  
	This condition states that the Ricci tensor should induce a focusing effect on timelike geodesic congruences. Mathematically, it requires that \( R_{\mu\nu}t^\mu t^\nu \geq 0 \) for any timelike vector \( t^\mu \) satisfying \( t^\mu t_\mu = -1 \). When applying Einstein's equations, this condition becomes 
	\[
	\left( T^{\text{eff}}_{\mu\nu} - \frac{1}{2}g_{\mu\nu}T \right)t^\mu t^\nu \geq 0.
	\]
	For a perfect fluid, this leads to the following inequalities
	\[
	\rho + p^{\text{eff}}\ \geq \ 0 \quad\quad \text{and} \quad\quad \rho + 3p^{\text{eff}}\ \geq \ 0
	\]
	applying the equation of state Eq.~\eqref{efergergre} on these conditions tends to 
	\begin{align}\label{dkjjwdbw}
		&
		F(\rho)\geq \frac{\sqrt{3}|\zeta|}{2\kappa}\rho+\frac{\rho^{3/2}}{\kappa}
		\\
		&F(\rho)\geq \frac{\sqrt{3}|\zeta|}{2\kappa}\rho
		\label{dkjwjdkw}
	\end{align} 
Fig. \ref{wdcwdcdw} illustrates the conditions given in Eqs.~\eqref{dkjjwdbw} and \eqref{dkjwjdkw}. The SEC is satisfied in the green region and violated in the red region.
\end{itemize}
\begin{figure}[h] 
	\centering 
	\includegraphics[width=3.6in]{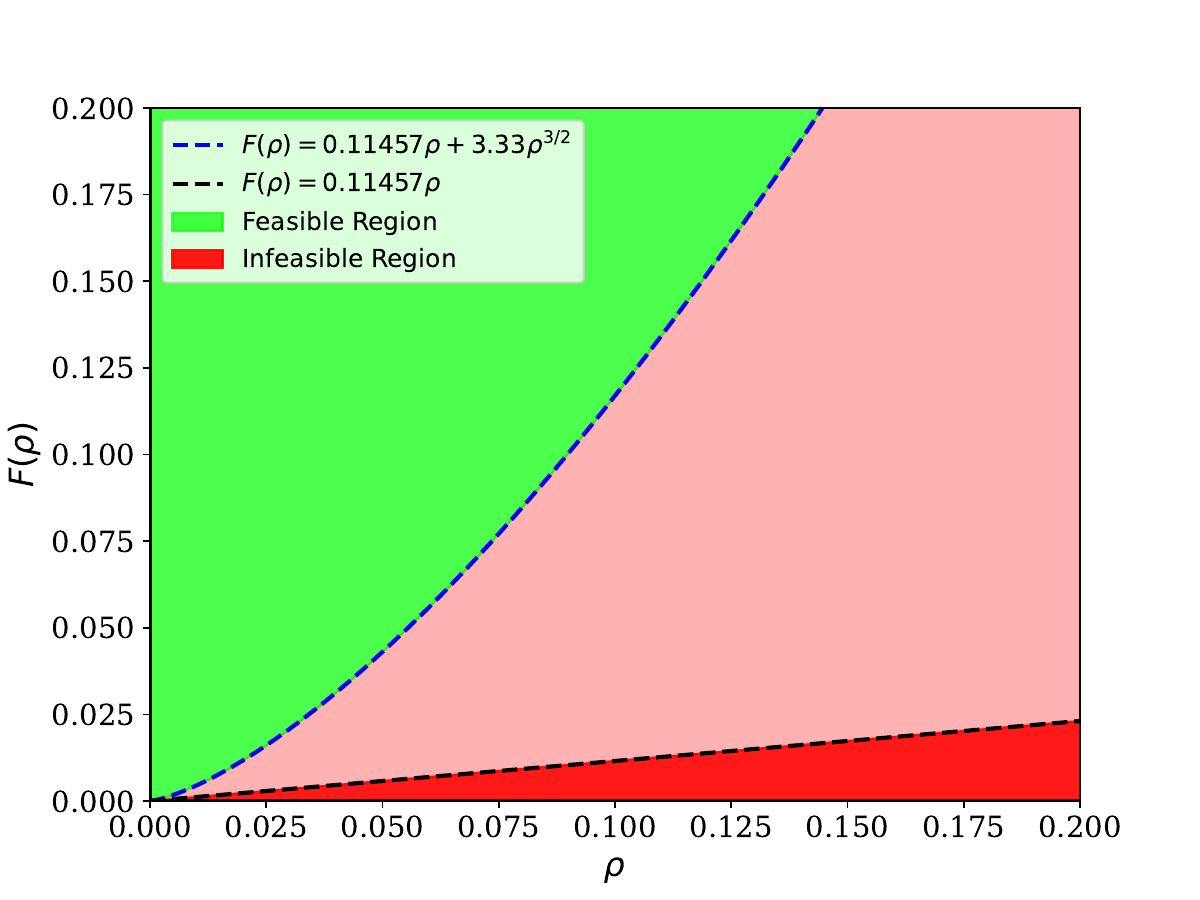}
	\caption{\label{wdcwdcdw}	
This figure illustrates the regions where the SEC is satisfied or violated. It is generated based on the conditions in Eqs.~\eqref{dkjjwdbw} and \eqref{dkjwjdkw} for 
 $\zeta=-0.04$ and $\kappa=0.3$. The red region indicates where both conditions are violated, the pink region corresponds to the violation of Eq.~\eqref{dkjwjdkw} while Eq.~\eqref{dkjjwdbw} is satisfied. The green region represents where both conditions hold.	}
\end{figure}

\section{Mori-Zwanzig projection operator formalism}\label{snwkdedef}
\subsubsection{Mori-Zwanzig Formalism}\label{csdjcadc}
For a variable $A$ that obeys Hamiltonian dynamics, the microscopic equation of motion is expressed as $\dot{A} = i\mathcal{L}A$, where $\mathcal{L}$ represents the Liouvillian operator.
Assuming that $\mathcal{L}$ is time-independent, one can express the microscopic equation of motion as $A(t) = e^{i\mathcal{L}t}A$, where $A(t)$ is the time-dependent observable in the Heisenberg picture, and $A$ is the time-independent observable in the Schrödinger picture. These definitions are fundamental for studying various microscopic systems. They can also be utilized to explore coarse-grained descriptions of a system. While several techniques exist for this purpose, the most powerful is the Mori-Zwanzig formalism \cite{Zwanzig:1960gvu, Mori:1965oqj}. This method is employed in non-equilibrium statistical mechanics to derive a generalized Langevin equation \cite{Rau:1995ea}. It characterizes the non-equilibrium behavior of functions defined within the phase space of a microscopic system, with a practical focus on coarse-grained variables.
The key aspect of the Mori-Zwanzig formalism is the definition of projection operators that differentiate between relevant and irrelevant variables. This approach facilitates the separation of a system's dynamics into relevant and irrelevant components, enabling the derivation of closed equations of motion for the relevant part.
To derive the Mori-Zwanzig equation, one must consider an arbitrary set of relevant variables $\{A_{j}\}$ and define the projection operator acting on a phase space $X$ as $
PX = A_{j}(A_{j}, A_{k})^{-1}(X, A_{k}),
$
where $(\ldots, \ldots)$ denotes a scalar product. Additionally, some mathematical tools are necessary, such as the Dyson identity, which is defined as follows
\begin{equation}\label{sdbjebdei}
e^{i\mathcal{L}t}=e^{iQ\mathcal{L}t}+\int^{t}_{0}dse^{i\mathcal{L}(t-s)}Pi\mathcal{L}e^{iQ\mathcal{L}s}
\end{equation}
where $Q = 1 - P$ is the orthogonal projection operator. Applying the Dyson identity to the expression $Qi\mathcal{L}A_{j}$ yields the Mori-Zwanzig equation, which is given by
\begin{equation}\label{sddbedbejw}
\dot{A}_{i}(t)=\Omega_{ij}A_{j}(t)+F_{i}(t)+\int^{t}_{0}dsK_{ij}(s)A_{j}(t-s)
\end{equation}	
where the functions defined in the above equation are as follows: the frequency matrix $ \Omega_{ij} = (A_{i}, A_{j})^{-1}(i\mathcal{L}A_{i}, A_{k}) $, the memory matrix $ K_{ij}(t) = (A_{i}, A_{j})^{-1}[i\mathcal{L}F_{i}(t), A_{k}] $, and the random force $ F_{i}(t) = e^{iQ\mathcal{L}t}Qi\mathcal{L}A_{i} $.
 The Mori-Zwanzig equation includes a memory term, indicating that the present rate of change $ \dot{A}_{i} $ depends not only on the current values of the variables but also on their values $ A_{i}(s) $ at previous times $ s < t $. This dependence is captured by the term $ K_{ij}(s) $ \cite{te2020projection}. 
\subsubsection{Application of the Mori-Zwanzig Formalism to the Buchert Equation}
\begin{figure}[h] 
	\centering 
	\includegraphics[width=3.6in]{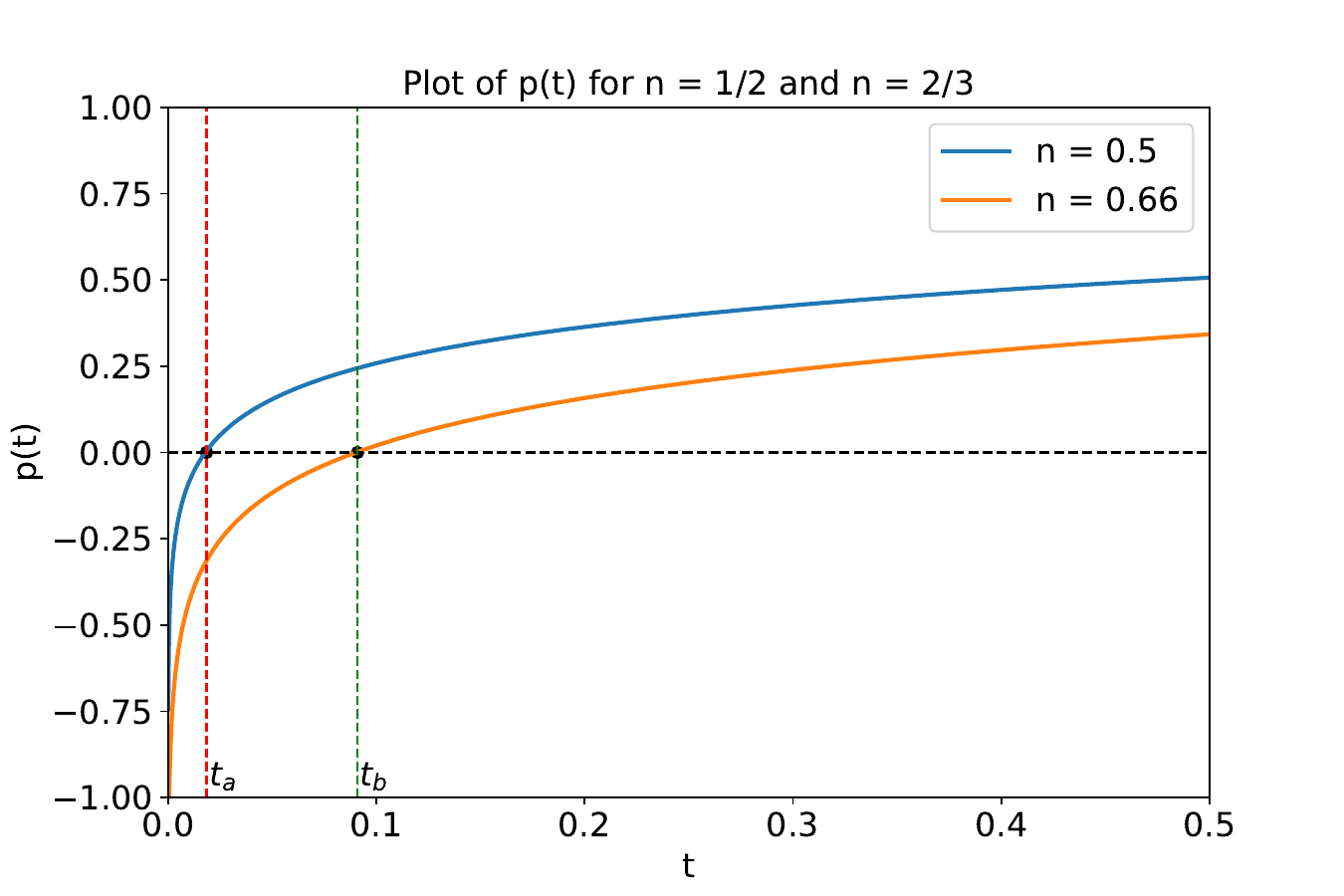}
	\caption{\label{figppp}
	This plot illustrates the behavior of pressure in Eq.~\eqref{wedj2edwei}. At small time values, the pressure is negative, while at large time values, it becomes positive.
	}
\end{figure}

The standard Friedmann equations apply to homogeneous and isotropic space-time, which aligns with experimental data at large scales with high accuracy. However, at intermediate scales, the effects of inhomogeneity must be considered. To address this, the Buchert equations have been derived \cite{Buchert:2001sa,Buchert:1999er,Buchert:2019mvq}, aiming to study an inhomogeneous universe through backreaction terms. \\
In a seminal paper \cite{Vrugt:2021sfu}, the authors employed the Mori-Zwanzig formalism to investigate the universe using the defined averaged parameters and the effects of backreaction terms. They considered a universe filled with irrotational dust matter and utilized the Buchert equations with an averaged pressure of $\bar{p}_{D} = 0$. In this paper, We aim to study early cosmic inflation, which necessitates accounting for the pressure term. The Buchert equations \cite{Buchert:2001sa, Buchert:2019mvq} with a general pressure function are given by
\begin{align}\label{dwejdbwe}
&\dot{H}+H^2=-\frac{(\bar{\rho}_{D}+3\bar{p}_D)}{6}+\frac{Q_{D}+\mathcal{P}_{D}}{3}
\\
&H^2=\frac{\bar{\rho}_{D}}{3}-\frac{\bar{R}_{D}+Q_{D}}{6}
\label{jjdndwkwp}
\end{align}
where $ \bar{\rho}_{D} $ is the averaged mass density, $ \bar{R}_{D} $ is the averaged spatial Ricci scalar, $ \mathcal{P}_{D} $ represents the dynamical backreaction, and $ Q_{D} $ denotes the kinematical backreaction term, which arises from the averaged fluctuations in the local expansion and shear scalars. Furthermore, there exists a differential relation between $ Q_{D} $, $ \mathcal{P}_{D} $, and $ \bar{R}_{D} $ \cite{Buchert:2001sa}, which can be expressed as follows
\begin{align}\label{dkcwdjkbo} \partial_{t}Q_{D}&+6HQ_{D}+\partial_{t}\bar{R}_{D}+2H\bar{R}_{D}+4H\mathcal{P}_{D}\notag\\ &-2\big[\dot{\bar{\rho}}_{D}+3H(\bar{\rho}_{D}+\bar{p}_{D})\big]=0
\end{align}
consider a situation where $ \dot{\bar{\rho}}_{D} + 3H(\bar{\rho}_{D} + \bar{p}_{D}) = 0 $ (which resembles the continuity equation for a perfect fluid), Eq.~\eqref{dkcwdjkbo} can be expressed as 
\[
\partial_{t}\bar{R}_{D} +\partial_{t}Q_{D} 
+2H(\bar{R}_{D} + 2\mathcal{P}_{D} + 3Q_{D}) = 0,
\]
which describes a differential relation among the backreaction terms and the averaged spatial Ricci scalar. \\
It is straightforward to combine the Buchert equations to formulate a single equation that does not depend on the averaged energy density,
\begin{equation}\label{djediede}
\dot{H}=-\frac{3}{2}H^{2}+\Psi	
\end{equation}
where $\Psi=Q_{D}/4-\bar{R}_{D}/12+\mathcal{P}_{D}/3-\bar{p}_{D}/2$.\\
To apply the Mori-Zwanzig formalism to the microscopic equation of motion Eq.~\eqref{djediede}, we utilize the Dyson identity Eq.~\eqref{sdbjebdei}. The resulting expression is as follows
\begin{align}\label{nnej}
\dot{H}(t)=&-\frac{3H^2(t)}{2}+\Omega_{HH} H(t)+\Omega_{HH^2} H^{2}(t)+F_{H}(t)\notag\\
&+\int_{0}^{t} ds\big[K_{HH}(s)H(t-s)+K_{HH^2}(s)H^2(t-s)\big]
\end{align}	
to arrive at the above equation, the relevant variables are taken to be $ A_{j} \in \{ H, H^2 \}$.
\\
It is well-established that the Buchert equations have been derived for the effective expansion factor $a_{D}(t)$ \cite{Buchert:2001sa, Buchert:1999er}, and the effective FRW metric has also been defined \cite{Desgrange:2019npu, Larena:2008be} as follows
\begin{equation}\label{dkbdjbd}
ds^{2}=-dt^2+L_{HD_{0}}^2a_{D}^2(t)\Big(\frac{dr_{D}^2}{1-k_{D}(t)r_{D}^2}+r_{D}^2d\Omega^2\Big)
\end{equation}
where $ L_{HD_{0}} = H^{-1}_{D_{0}} a^{-1}_{D_{0}} $, and $ k_{D} $ is a dimensionless spatial constant-curvature function related to the averaged spatial Ricci scalar through 
\begin{equation}\label{dkjdcjwd}
k_{D}(t) = \frac{\bar{R}_{D}(t)}{|\bar{R}_{D_{0}}|} \frac{a^{2}_{D}(t)}{a^{2}_{D_{0}}}.
\end{equation}
For spatially flat space-time, where $k_{D}=\bar{R}_{D}(t)= 0$, we consider $L_{HD_{0}} = 1$ and redefine $a_{D}(t) = a(t)$. \\
By substituting the spatially flat effective FRW metric Eq.~\eqref{dkbdjbd} into Einstein's field equations $G_{\mu\nu} = T^{\text{eff}}_{\mu\nu}$, along with the effective perfect fluid  energy-momentum tensor 
\begin{equation}\label{jwedjeedj}
T^{\text{eff}}_{\mu\nu} = \big(\rho(t) + p^{\text{eff}}(t)\big) u_{\mu} u_{\nu} + p^{\text{eff}}(t) g_{\mu\nu},
\end{equation}
we obtain the following Friedmann equations
\begin{align}\label{dndeenf}
&H^2(t)=\frac{\rho(t)}{3}
\\
&\dot{H}(t)=-\frac{\rho(t)+p^{\text{eff}}(t)}{2}
\label{djedkcjbedjb}.
\end{align}	
Comparing Eq.~\eqref{nnej} with Eq.~\eqref{dndeenf} leads to
\begin{align}\label{dhbwidbweo}
	p^{\text{eff}}(t)=&\underbrace{-2\big(\Omega_{HH} H(t)+\Omega_{HH^2} H^{2}(t)+F_{H}(t)\big)}_{\text{oblivion term}}
	\notag\\
	&\underbrace{-2\int_{0}^{t} ds\big[K_{HH}(s)H(t-s)+K_{HH^2}(s)H^2(t-s)\big]}_{\text{memory term}}
\end{align}	
The full form of the above equation is difficult to handle, but it can be simplified under specific conditions by neglecting certain terms. In particular, this is done by disregarding the first line on the right-hand side of Eq.~\eqref{dhbwidbweo}. Additionally, further simplifications are introduced (see Ref.~\cite{Vrugt:2021sfu}), namely $K_{HH^2} = 0$ and $K_{HH} = -\kappa e^{-\zeta t}$. As a result, the pressure Eq.~\eqref{dhbwidbweo} takes the form
\begin{equation}\label{dndddjw}
p^{\text{eff}}(t)=2\kappa \int_{0}^{t} dse^{-\zeta s}H(t-s)
\end{equation}
This equation describes a memory effect in pressure, where the present value of $ p^{\text{eff}}(t)$ depends on its past values at earlier times $ s < t$, specifically governed by the previous values of the Hubble parameter. 
Determining an explicit function for the pressure without specifying the Hubble parameter $H(t-s)$ in the integral proves to be challenging. However, it can be formulated as a differential equation
\begin{equation}\label{djbwedwo}
\dot{p}^{\text{eff}}+\zeta p^{\text{eff}}-2\kappa H(t)=0.	
\end{equation} 
In the case where the Hubble parameter remains nearly constant, resembling early cosmic inflation with $H(t) \approx \alpha$, the differential equation \eqref{djbwedwo} can be solved as
\begin{equation}\label{dwjkw}
p^{\text{eff}}(t)=\frac{2\kappa \alpha}{\zeta}	+\beta e^{-\zeta t}
\end{equation}
where $\beta$ is an integral constant. It is straightforward to rewrite the above equation in terms of the effective scale factor $a(t)$. Using the definition of the Hubble parameter, we obtain $a(t) = a_{0} e^{\alpha t}$. Substituting this into Eq.~\eqref{dwjkw} yields
\begin{equation}\label{ddjnwjeedd}
 p^{\text{eff}}=\frac{2\kappa\alpha}{\zeta}+\beta \big(\frac{a(t)}{a_{0}}\big)^{\frac{-\zeta}{\alpha}}.
 \end{equation}
For negative values of $\zeta = -|\zeta|$, Eq.~\eqref{ddjnwjeedd} takes the form
$
p^{\text{eff}} = -2\kappa\alpha/|\zeta| + \beta\left(a(t)/a_{0}\right)^{|\zeta|/\alpha}.
$
Thus, in the limit $a(t) \to 0$, the pressure approaches a constant, finite value. 
Furthermore, substituting Eq.~\eqref{ddjnwjeedd} into the continuity equation gives
\begin{equation}\label{dkcwdbfiuwdb}
\rho = -\frac{3\alpha\beta}{|\zeta| + 3\alpha}\left(\frac{a(t)}{a_{0}}\right)^{\frac{|\zeta|}{\alpha}} + \frac{2\kappa\alpha}{|\zeta|}.
\end{equation}
At small distances, this leads to the energy density approaching $\rho|_{a(t) \to 0} \simeq 2\kappa\alpha/|\zeta|$, with the state parameter given by $\omega \approx -1$. 
Combining the energy density Eq.~\eqref{dkcwdbfiuwdb} with the pressure Eq.~\eqref{ddjnwjeedd}, we obtain the equation of state
\begin{equation}\label{dlcnwdwd}
	\omega(\rho) = \frac{p^{\text{eff}}}{\rho} = -\left(1+ \frac{|\zeta|}{3\alpha}\right) + \frac{2\kappa}{3\rho},
\end{equation}
where, in the high-energy density limit ($\rho \gg \kappa$ and $\alpha \gg |\zeta|$), the state parameter tends to $\omega \approx -1$.
\\
One can also explore other forms of the scale factor, such as $a(t) \propto t^{n}$, where $n = 1/2$ corresponds to the radiation-dominated era and $n = 2/3$ represents the matter-dominated era. For this choice, the Hubble parameter is given by $H(t) = n/t$. Substituting this into Eq.~\eqref{djbwedwo}, the pressure equation takes the form
\begin{equation}\label{wedj2edwei}
	p^{\text{eff}}(t) = \kappa n e^{-\zeta t} \operatorname{Ei}(\zeta t) + C_{1} e^{-\zeta t},
\end{equation}
where $C_{1}$ is an integration constant, and the exponential integral function is defined in Appendix \ref{wckjwcdwdn}. The pressure described in Eq.~\eqref{wedj2edwei} is illustrated in Fig.~\ref{figppp}, which demonstrates that the pressure is negative for small time values and transitions to positive for larger time values.
 \\
\\
Another significant feature of Eq.~\eqref{djbwedwo} is that substituting the Friedmann equation \eqref{dndeenf} along with the continuity equation $ p^{\text{eff}}(t) = -\dot{\rho}(t)/3H(t) - \rho(t)$ into Eq.~\eqref{djbwedwo} yields the following differential equation
\begin{equation}\label{dwjkebe}
\ddot{H}(t)+3H\dot{H}(t)+\zeta\big(\dot{H}(t)+\frac{3}{2}H^2(t)\big)+\kappa H(t)=0
\end{equation}	
This corresponds to the equation derived in \cite{Vrugt:2021sfu}. We employed the Einstein field equations for the effective FRW metric and considered the energy-momentum tensor as a perfect fluid, ultimately leading to the same equation obtained through an alternative approach (see Appendix \ref{sdqdnwd}).

\section{Cosmological Inflation In Mori-Zwanzig Formalism}\label{nbsdibwdw}
We aim to apply the conditions of early cosmic inflation to Eq.~\eqref{dwjkebe}. Within this framework, the slow-roll parameters described in Eq.~\eqref{djndcjw} are defined in relation to the Hubble parameter as follows
\begin{align}\label{dcnwdbwj}
&\epsilon_{1}=-\frac{\dot{H}(t)}{H^2(t)}	\ll 1, \\
&\epsilon_{2}=\frac{\ddot{H}(t)}{\dot{H}(t)H(t)}-2\frac{\dot{H}(t)}{H^2(t)}\ll1
\label{dcwdldcwin}
\end{align}
considering the above conditions one can write Eq.~\eqref{dwjkebe} in the  form 
\begin{equation}\label{dwcdjnwdonw}
\dot{H}(t)+\frac{\zeta}{2} H(t)+\frac{\kappa}{3}\simeq0
\end{equation}
this is indeed the Friedmann equation \eqref{dcdlkwddlwnd}. It can be straightforwardly rewritten in terms of the energy density and pressure with equation of state
\begin{equation}
\label{cwdcnwdcd}
\omega(\rho)\simeq-1+{\frac{\zeta}{\sqrt{3\rho}}}+\frac{2\kappa}{3\rho}
\end{equation}
at high energy densities $\rho \gg \{\zeta^2, \kappa\}
$, the state parameter approaches $\omega \approx -1$.

\begin{figure}[h] 
	\centering 
	\includegraphics[width=3.7in]{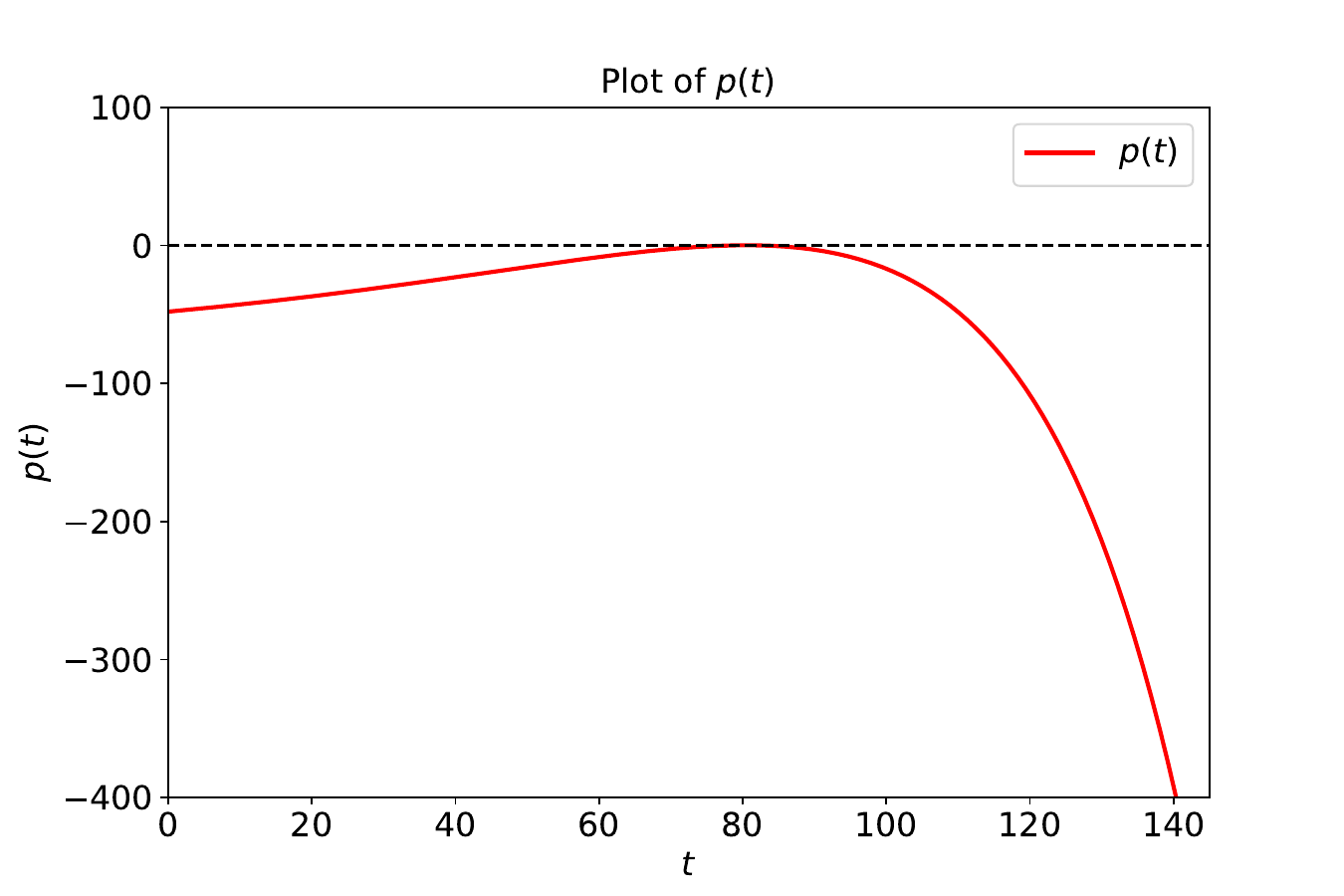}
	\caption{\label{refgefg}
This plot illustrates the behavior of the pressure equation \eqref{ddfwdfwdf}. It shows that the pressure remains negative for all time values, decreases to zero at early times, and increases at later times.
	}
\end{figure}

\begin{figure}[h] 
	\centering 
	\includegraphics[width=3.7in]{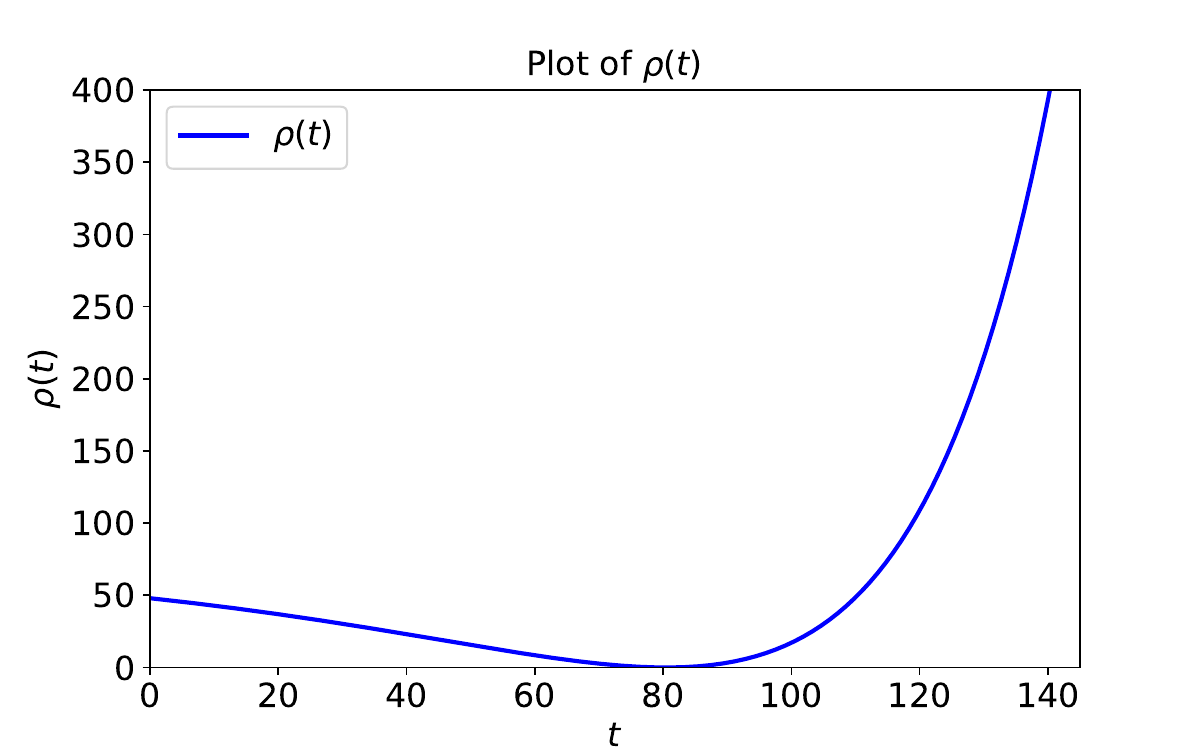}
	\caption{\label{figgg}	
		This plot depicts the behavior of the energy density given by Eq.~\eqref{dfnedfninfe}, showing that it remains positive for all time values, decreases to zero at early times, and increases at later times.
	}
\end{figure}

One can write this equation of state in a form similar to the dynamical dark energy (DDE) parametrization (CPL model see Ref.~\cite{Chevallier:2000qy,Linder:2002et}), 
$
\omega(a)=\omega_{0}+\omega_{a}(1-a),
$
which is commonly used to describe the late–time universe. 
However, in order to apply such a parametrization consistently to the early universe, 
we propose a modified form
$
\omega(a)=\omega_{0}+\omega_{a}(a_{i}-a),
$
where $a_{i}$ is not the present scale factor, but rather an initial condition that can be chosen at any epoch, ranging from the early to the late universe. 
In this way, Eq.~\eqref{cwdcnwdcd} can be rewritten in the form
\begin{equation}\label{wdcwde}
	\begin{aligned}
		\omega(a) &= \omega_0 + \omega_a (a_i - a), \\
		\omega_0 &= -1 + \frac{\zeta}{\sqrt{3\rho_i}} + \frac{2\kappa}{3\rho_i}, \\
		\omega_a &= -\frac{3\,\delta_i\,\rho_i}{a_i} 
		\left( \frac{\zeta}{2\sqrt{3}\,\rho_i^{3/2}} + \frac{2\kappa}{3\,\rho_i^{2}} \right)
	\end{aligned}
\end{equation}
The details of the calculations leading to Eq.~\eqref{wdcwde} are given in Appendix~\ref{dmdwdcw}.\\
Recent analyses of experimental data suggest that $\omega_0 > -1$ and $\omega_a < 0$ (see Ref.~\cite{DESI:2025zgx,DESI:2024mwx}).
\\
The equation \eqref{dwcdjnwdonw} is a differential equation for the Hubble parameter, and its solution is given by
\begin{equation}\label{Heq}
	H(t)=-\frac{2 \kappa}{3 \zeta}+\gamma \exp\Big(-\frac{\zeta}{2}t\Big),
\end{equation}
where $\gamma$ is the integration constant. 
One can derive the scale factor as a function of time from the Hubble parameter Eq.~ \eqref{Heq}, which is expressed as follows
\begin{equation}\label{scale factor}
a(t)=a_{0} \exp\bigg(-\frac{2 \kappa t+6 \gamma\exp\left(-\zeta t/2\right)}{3 \zeta}\bigg).
\end{equation}
Furthermore, the time derivative of the Hubble parameter can be derived from the previously discussed relations, expressed as
\begin{equation}\label{Hdot}
	\dot{H}=-\frac{\gamma \zeta}{2}\exp\Big(-\frac{\zeta }{2}t\Big).
\end{equation}
We will use this time derivative in the forthcoming calculations. Taking into account the spatially flat effective FLRW spacetime given by Eq.~\eqref{dkbdjbd} and assuming that the universe is dominated by a perfect fluid with energy density 
$\rho$ and pressure $p^{\text{eff}}$, we can derive the Friedmann equations for the model as follows, using Eqs.~\eqref{Heq} and \eqref{Hdot}
\begin{align}
\label{dfnedfninfe}
&3H^2\equiv \rho(t)=3\bigg(\gamma \exp \Big(-\frac{\zeta }{2}t\Big)-\frac{2\kappa}{3\zeta}\bigg)^2,\\
&3H^2+2\dot{H}\equiv -p^{\text{eff}}(t)=-\gamma \zeta \exp \Big(-\frac{\zeta }{2}t\Big)\nonumber\\
&\hspace*{3cm}+3\bigg(\gamma \exp \Big(-\frac{\zeta }{2}t\Big)-\frac{2\kappa}{3\zeta}\bigg)^2.
\label{ddfwdfwdf}
\end{align}
The evolution of the energy density described by Eq.~\eqref{dfnedfninfe} over time is depicted in Fig.\ref{refgefg}, whereas the pressure characterized by Eq.~\eqref{ddfwdfwdf} is illustrated in Fig.\ref{figgg}.

\begin{figure}[htbp!] 
	\centering 
	\includegraphics[width=3.8in]{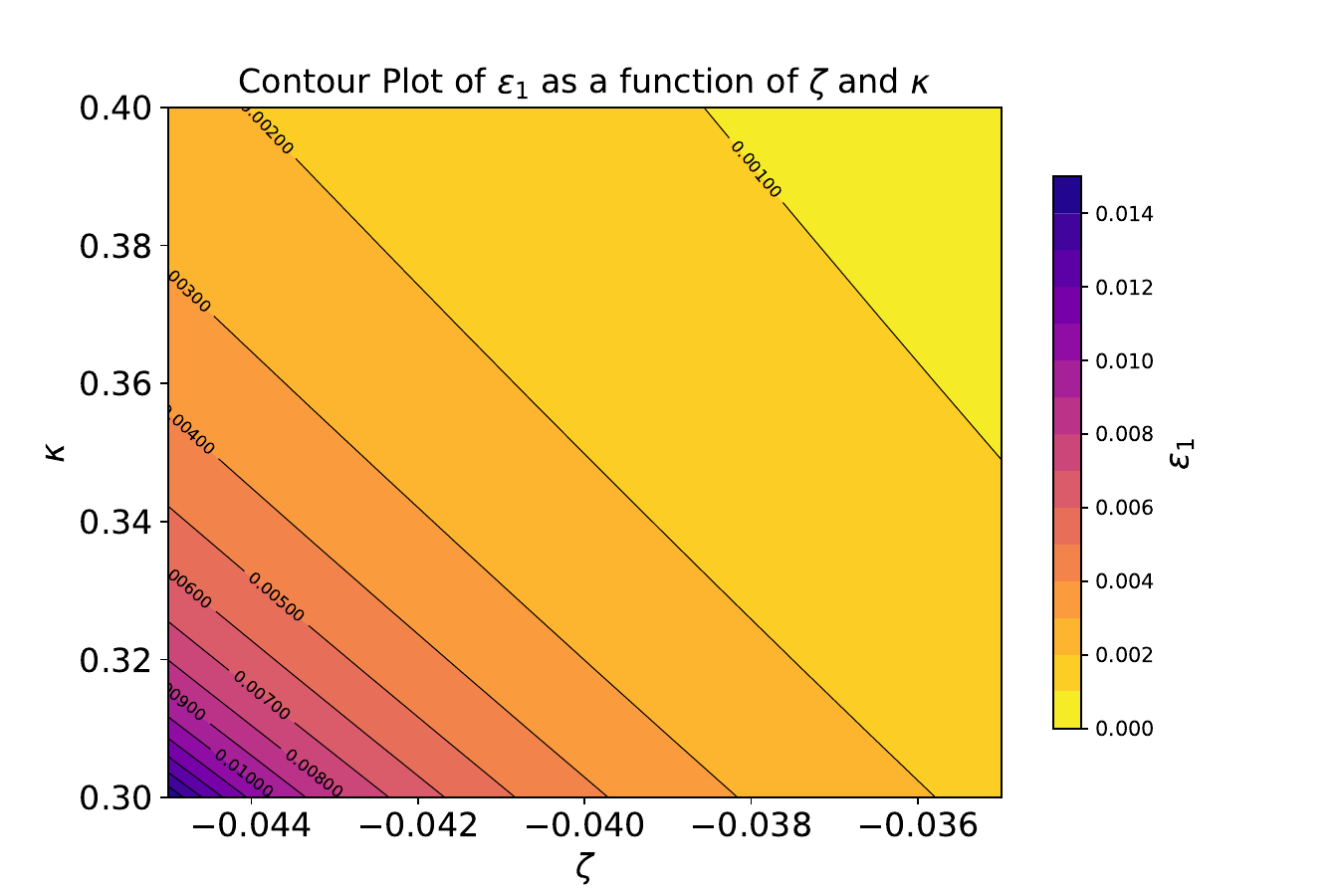}
	\caption{\label{fig5}
		Contour plot of the first slow-roll parameter $\epsilon_1$ as a function of the model parameters $\kappa$ and $\zeta$. The horizontal axis represents $\kappa$ in the range $0.30 \leq \kappa \leq 0.40$, while the vertical axis corresponds to $\zeta$ spanning $-0.045 \leq \zeta \leq -0.035$. The contour lines indicate regions of constant $\epsilon_1$, with variations in shading representing different magnitudes.	
	}
\end{figure}

\begin{figure}[htbp!] 
	\centering 
	\includegraphics[width=3.8in]{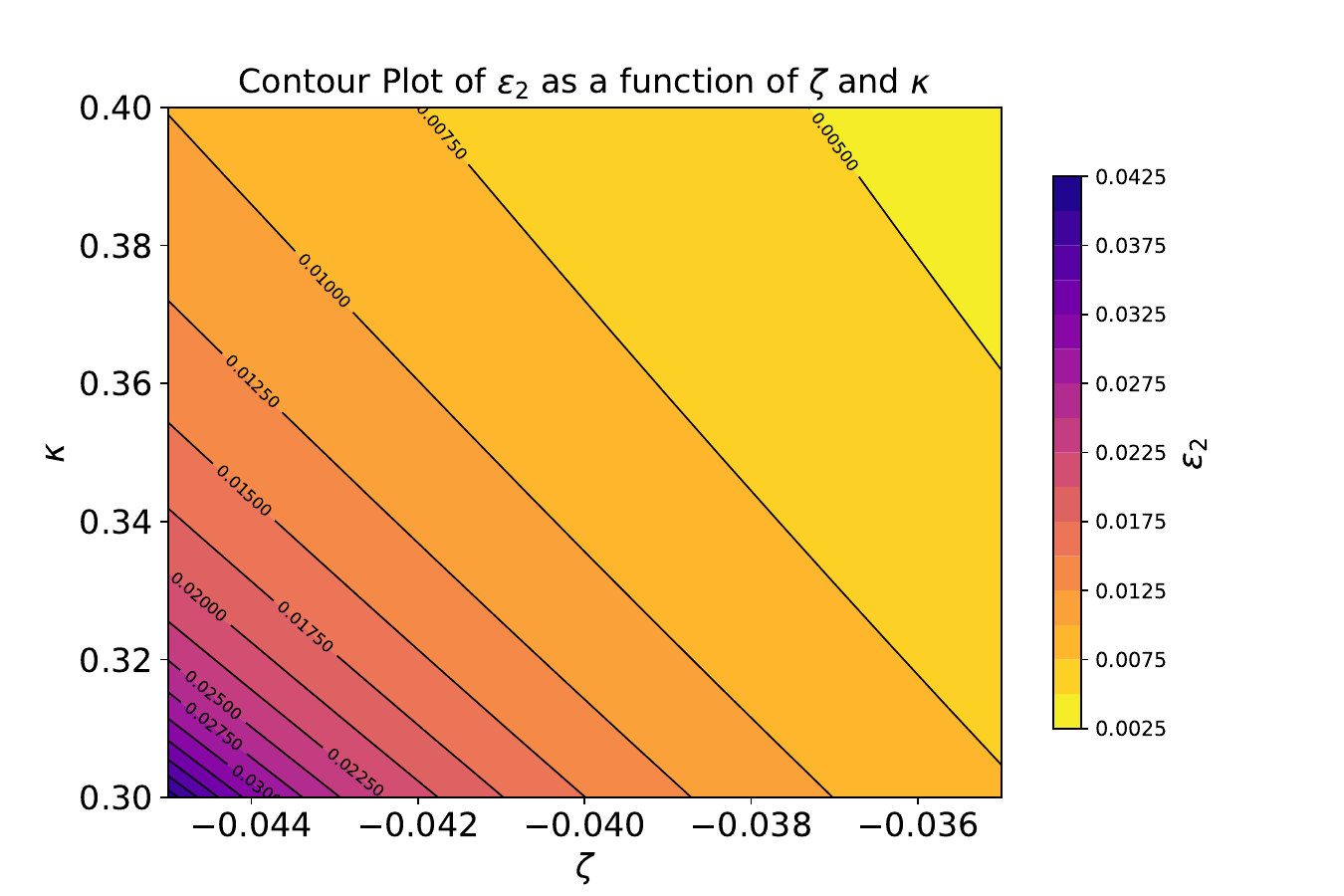}
	\caption{\label{fig6}	
		The contour plot illustrates the second slow-roll parameter $\epsilon_2$ as a function of the model parameters $\kappa$ and $\zeta$. The horizontal axis spans $0.30 \leq \kappa \leq 0.40$, while the vertical axis covers the range $-0.045 \leq \zeta \leq -0.035$. Contour lines represent constant values of $\epsilon_2$, with shading variations indicating different magnitudes.	}
\end{figure}

\begin{figure}[htbp!] 
	\centering 
	\includegraphics[width=3.8in]{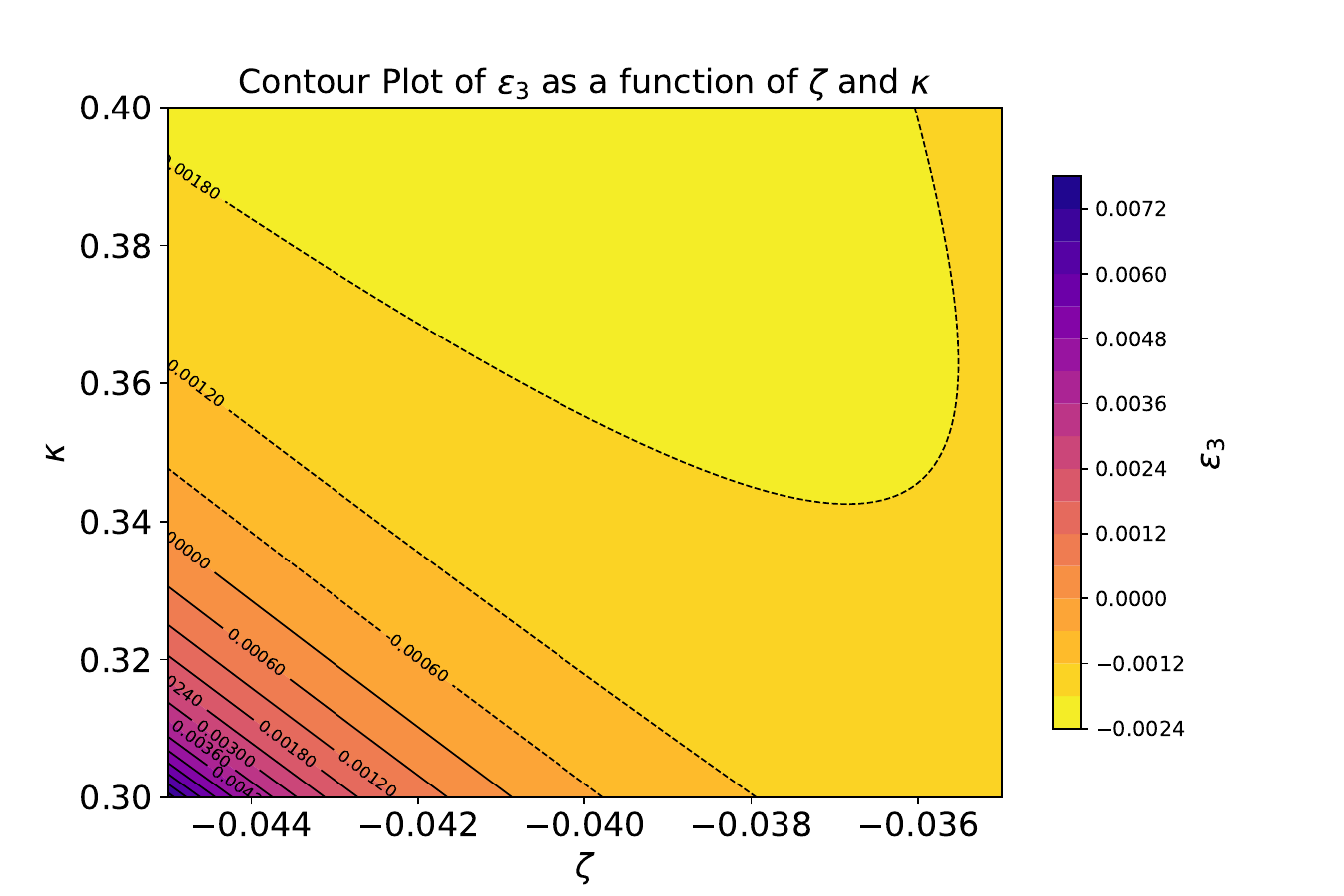}
	\caption{\label{fig7}
			The contour plot depicts the third slow-roll parameter $\epsilon_3$ as a function of the model parameters $\kappa$ and $\zeta$. On the horizontal axis, $\kappa$ is represented within the range $0.30 \leq \kappa \leq 0.40$, while the vertical axis corresponds to $\zeta$, which spans $-0.045 \leq \zeta \leq -0.035$. The contour lines illustrate regions of constant $\epsilon_3$, with differences in shading indicating various magnitudes.	
		}
\end{figure}

\begin{figure}[htbp!] 
	\centering 
	\includegraphics[width=3.7in]{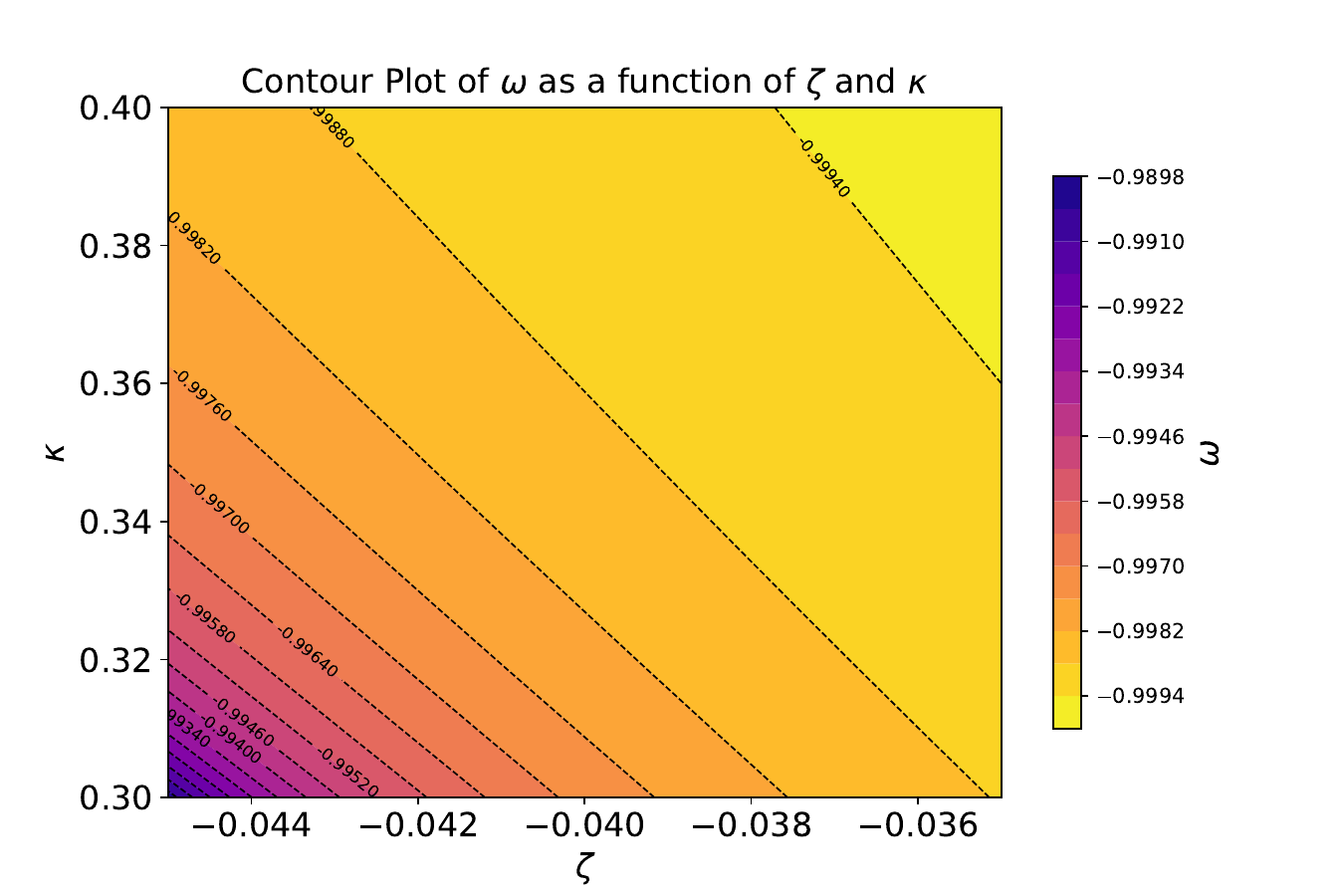}
	\caption{\label{fig8}
	The contour plot displays the equation of state parameter $\omega$ as a function of the model parameters $\kappa$ and $\zeta$. The horizontal axis shows $\kappa \in [0.30, 0.40]$, while the vertical axis spans $\zeta \in [-0.045, -0.035]$. Contour lines indicate constant values of $\omega$, and the shading variations represent different magnitudes. The plot suggests that $\omega \approx -1$.
	}
\end{figure}

Furthermore, considering the equation of state $p = w\rho$, the state parameter can be derived as follows
\begin{align}\label{EOS}
	w(t)=-1-\frac{2}{3}\frac{\dot{H}}{H^2}=-1+\frac{3\gamma\zeta^3 \exp\big(-\frac{\zeta}{2}t\big)}{\Big(2\kappa-3\gamma\zeta \exp\big(-\frac{\zeta}{2}t\big)\Big)^2}.
\end{align}
This equation demonstrates that when $H=0$, the state parameter diverges as $\omega \to \infty$. This divergence occurs at a specific time given by $t_{0}=-(2/\zeta)\ln(2\kappa/3\gamma \zeta)$. Furthermore, Eq.~\eqref{EOS} indicates that under the slow-roll condition $\epsilon_{1} \ll 1$, the state parameter approaches $\omega \simeq -1$. 
\\
We can now apply Eq.~\eqref{Heq} to investigate inflation within the Mori-Zwanzig formalism. To accomplish this, we first define the Hubble slow-roll parameters $\epsilon_i$ for 
$i = \{1, 2, 3\}$ (as outlined in Appendix \ref{dsbdddjowbd}). It is known that if the condition $|\epsilon_i| \ll 1$ is satisfied, inflation will occur and persist long enough o address the standard cosmological problems. Moreover, inflation ends when the first slow-roll parameter reaches one, i.e., \(\epsilon_1 = 1\). The slow-roll parameters can be expressed in terms of the e-folding number, which quantifies the rate of expansion during inflation as the natural logarithm of the scale factor, as discussed in \cite{Liddle:1999mq}
\begin{equation}\label{Ndef}
	N\equiv \ln\left(\frac{a_{\rm end}}{a}\right)=\int_t^{t_{\rm end}} H {\rm d}t,
\end{equation}
where the index `end` denotes the values of quantities at the end of inflation. Therefore, using Eq.~\eqref{Ndef}, we obtain the relation ${\rm d}/{\rm d}t = H {\rm d}/{\rm d}N$.
Additionally, Eq.~\eqref{EOS} can be rewritten as
\begin{equation}\label{EOSN}
	w=-1-\frac{2}{3}\frac{H^\prime(N)}{H(N)}.
\end{equation}
where the prime symbol denotes the derivative with respect to $N$. Considering Eq.~\eqref{eps1N} in conjunction with the condition $\epsilon_1 \ll 1$, it follows that the equation above strongly supports the idea that during the inflationary era, the equation of state parameter remains approximately $w \approx -1$, ensuring the accelerated expansion necessary to address key cosmological challenges.
 This result aligns with the predictions of Eqs.~\eqref{cwdcnwdcd} and \eqref{dkcwdbfiuwdb} in the high-energy density regime.\\
To express the e-folding number in terms of time, Eq.~\eqref{Heq} is substituted into Eq.~\eqref{Ndef}, resulting in the following expression
\begin{equation}\label{Nt}
	N=-\frac{2\kappa}{3\zeta} t-\frac{2\gamma}{\zeta} \exp\left(\frac{-\zeta}{2}t\right), 
\end{equation}
from the above relation, time can be readily expressed in terms of the e-folding number, given by
\begin{equation}\label{tN}
t=\frac{4 \kappa{\rm A(N)}-3 \zeta^2 N}{2\kappa \zeta}
\end{equation}
the function $A(N)$ is defined as
\begin{equation}\label{defA}
	{\rm A(N)}\equiv {\rm W}\bigg(\frac{-3\zeta \gamma\exp\big(3N\zeta^2/(4\kappa)\big)}{2\kappa}\bigg)
\end{equation}
where $W(x)$ denotes the Lambert function, as defined in Appendix \ref{rtergttr}. By substituting Eq.~\eqref{tN} into Eq.~\eqref{Heq}, we can express the Hubble parameter in terms of the e-folding number as follows
\begin{align}\label{HN}
H(N)=
-\frac{2\kappa}{3\zeta}\big(1+A(N)\big).
\end{align}
The derivatives of the Hubble parameter with respect to the e-folding number can be derived from Eq.~\eqref{HN} in the following way
\begin{align}
\label{H1p}
&H^\prime(N)=\frac{-\zeta A(N)}{2\big(1+A(N)\big)},\\
\label{H2p}
&H^{\prime\prime}(N)=\frac{-3\zeta^3 A(N)}{8\kappa\big(1+A(N)\big)^3},\\
\label{H3p} &H^{\prime\prime\prime}(N)=-\frac{9\zeta^5A(N)\big(1-2A(N)\big)}{32\kappa^2\big(1+A(N)\big)^5}.
\end{align}
Substituting Eqs.~\eqref{HN}, \eqref{H1p}, \eqref{H2p}, and \eqref{H3p} into Eqs.~\eqref{eps1N}, \eqref{eps2N}, and \eqref{eps3N}, the slow-roll parameters can be expressed as
\begin{align}
\label{eps1NN}
&\epsilon_1=-\frac{3\zeta^2A(N)}{4\kappa\big(1+A(N)\big)^2}
\\
&\epsilon_2=\frac{3\zeta^2\big(1-A(N)\big)}{4\kappa\big(1+A(N)\big)^2}
\\
&\epsilon_3=\frac{3\zeta^2\big(A^{2}(N)-2A(N)-1\big)}{4\kappa\big(1+A(N)\big)^2\big(1-A(N)\big)}
\end{align}
Moreover, substituting Eq.~\eqref{H1p} and Eq.~\eqref{HN} into Eq.~\eqref{EOSN} allows for a straightforward expression of the state parameter as a function of e-folding number
\begin{equation}\label{lastEOSN}
	w=-1-\frac{\zeta^2A(N)}{2\kappa\big(1+A(N)\big)^2}.
\end{equation}
The behavior of the slow-roll parameters \(\epsilon_1\), \(\epsilon_2\), and \(\epsilon_3\) is shown in Fig.~\ref{fig5}, Fig.~\ref{fig6}, and Fig.~\ref{fig7}, respectively, as functions of the model parameters \(\zeta\) and \(\kappa\), with \(\gamma = -1\) and \(N = 70\). The results for the intervals \(\zeta \in [-0.045, -0.035]\) and \(\kappa \in [0.3, 0.4]\) satisfy the condition \(\epsilon_i \ll 1\). Additionally, in Fig.~\ref{fig8}, the value of the equation of state parameter \(\omega\) has been determined for the given values and found to match the expected value for the inflationary era.

\begin{figure}[htbp!] 
	\centering 
	\includegraphics[width=3.7in]{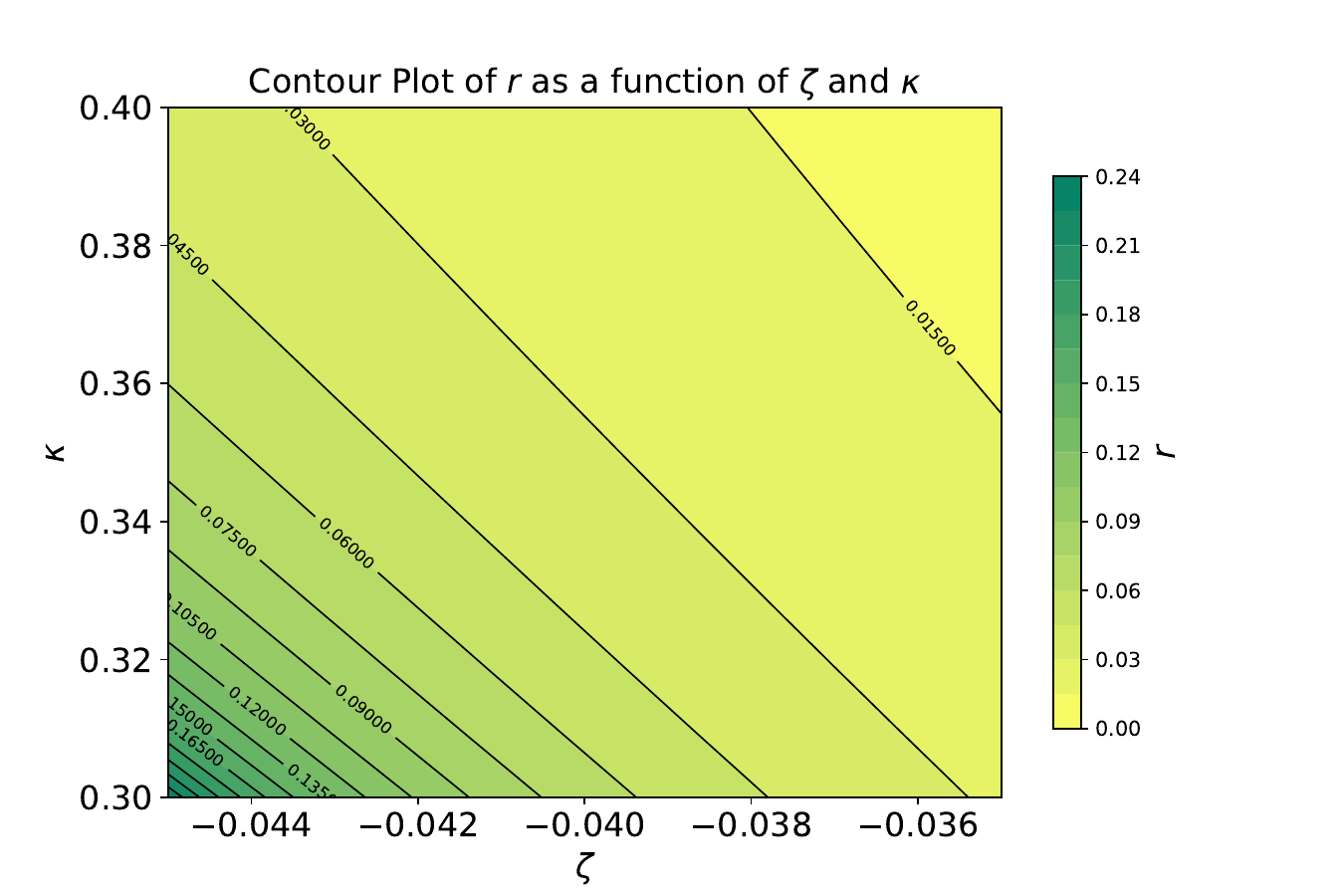}
	\caption{\label{figffh}	
	The contour plot illustrates the tensor-to-scalar ratio $r$ as a function of the model parameters $\kappa$ and $\zeta$. The horizontal axis represents $\kappa \in [0.30, 0.40]$, while the vertical axis covers $\zeta \in [-0.045, -0.035]$. Contour lines indicate constant values of $r$, with shading variations denoting different magnitudes.
	}
\end{figure}

\begin{figure}[htbp!] 
	\centering 
	\includegraphics[width=3.7in]{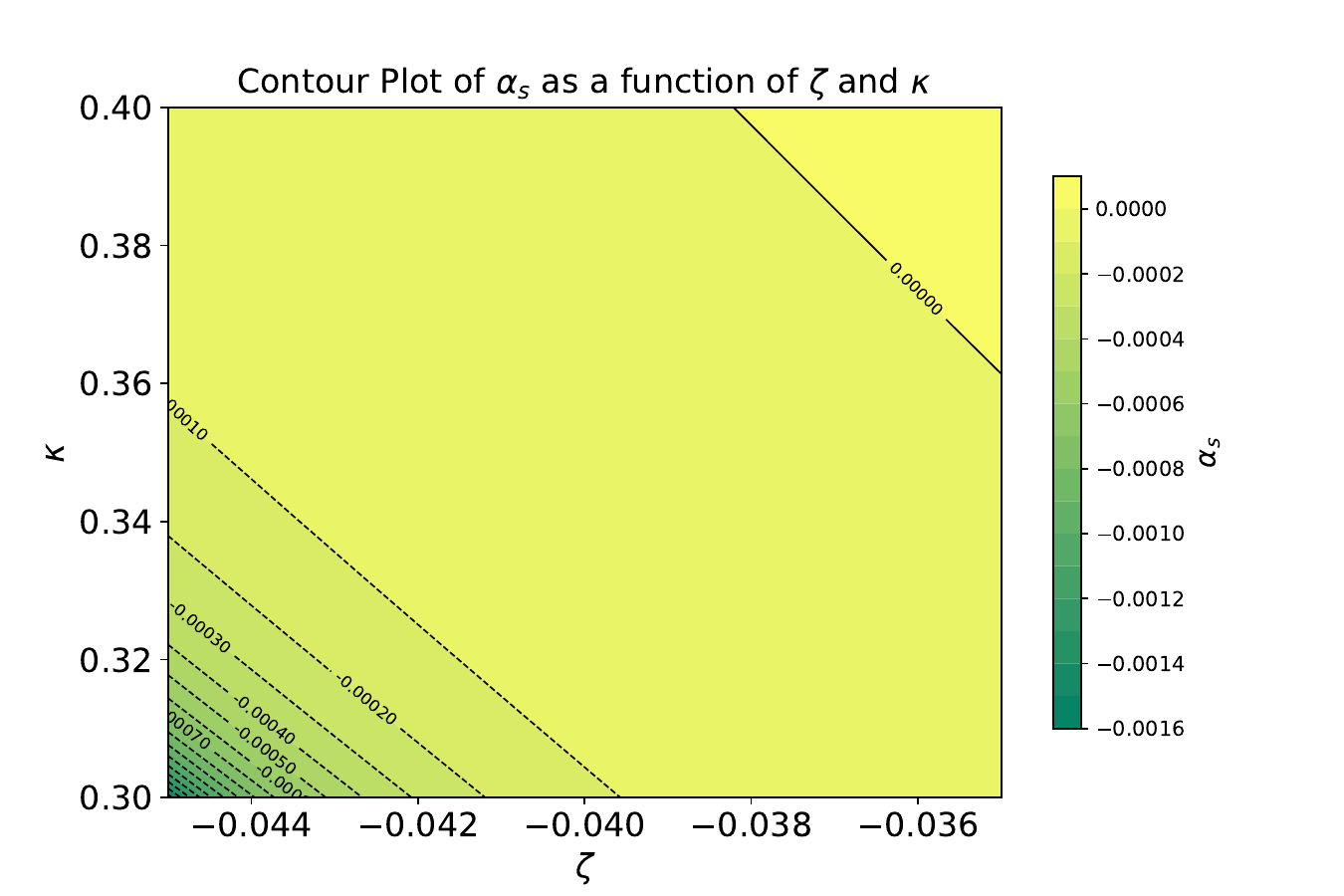}
	\caption{\label{figdd}
The contour plot displays the running of the scalar spectral index $\alpha_{s}$ as a function of the model parameters $\kappa$ and $\zeta$. The horizontal axis shows $\kappa \in [0.30, 0.40]$, while the vertical axis spans $\zeta \in [-0.045, -0.035]$. Contour lines indicate constant values of $\alpha_{s}$, and the shading variations represent different magnitudes.
	}
\end{figure}

\begin{figure}[htbp!] 
	\centering 
	\includegraphics[width=3.7in]{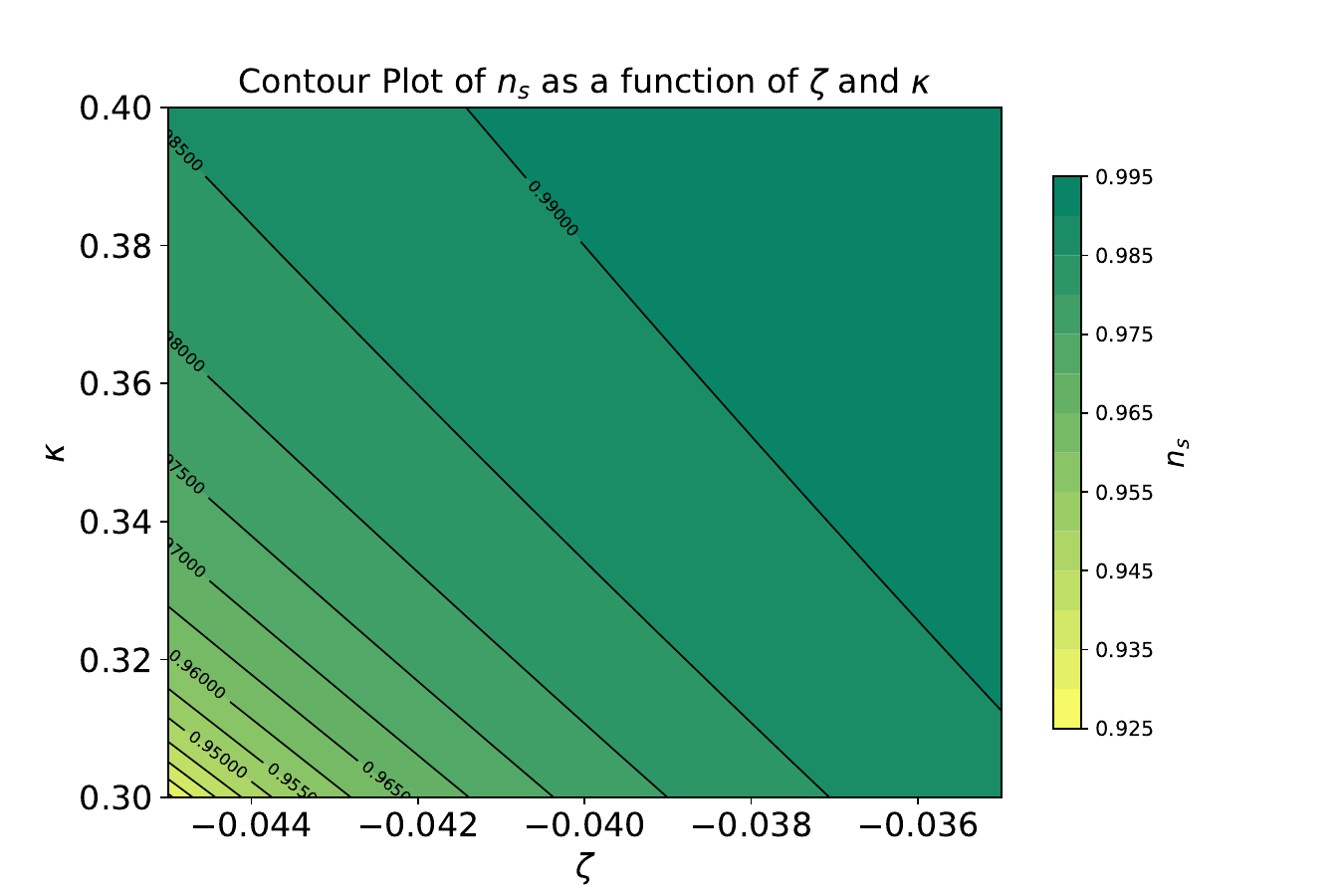}
	\caption{\label{fighh}
	The contour plot illustrates the running of the scalar spectral index $n_{s}$ as a function of the model parameters $\kappa$ and $\zeta$. The horizontal axis covers the range $\kappa \in [0.30, 0.40]$, while the vertical axis spans $\zeta \in [-0.045, -0.035]$. Contour lines denote constant values of $n_{s}$, with shading variations indicating different magnitudes.
	}
\end{figure}
\begin{figure}[htbp!] 
	\centering 
	\includegraphics[width=3.7in]{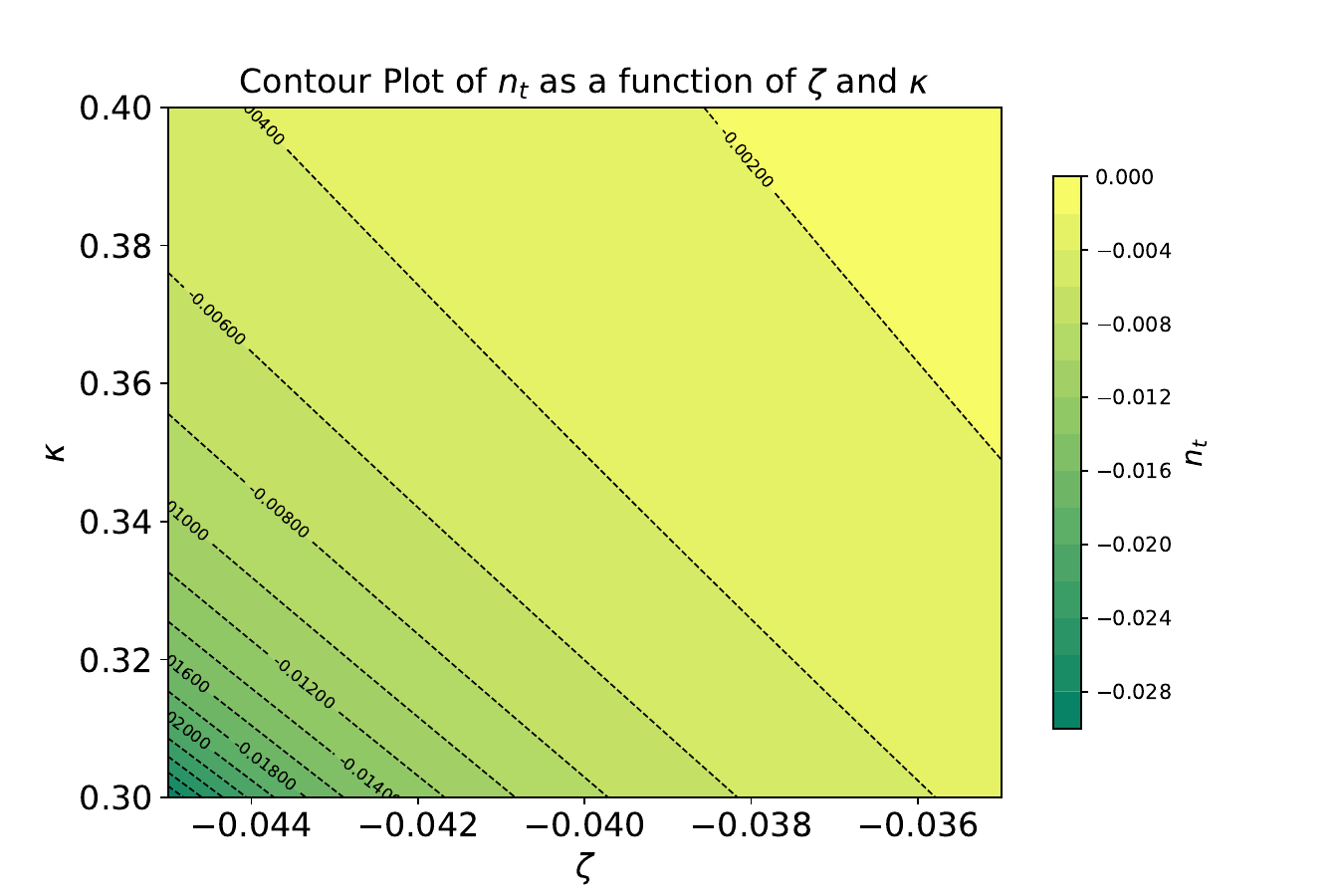}
	\caption{\label{figh}
		The contour plot illustrates the running of the tensor spectral index $n_{t}$ as a function of the model parameters $\kappa$ and $\zeta$. The horizontal axis covers $\kappa \in [0.30, 0.40]$, while the vertical axis spans $\zeta \in [-0.045, -0.035]$. Contour lines represent constant values of $n_{t}$, with shading variations indicating different magnitudes.	
	}
\end{figure}
The main objective of gravitational theories in describing cosmological inflation is to ensure that their predictions match observational data. To accomplish this, the theoretical predictions of the proposed gravitational models must be evaluated for the following inflationary observables: the scalar spectral index $n_{\rm s}$, the tensor spectral index $n_{\rm t}$, the tensor-to-scalar ratio $r$, and the running of the scalar spectral index $\alpha_{\rm s}$.\\
In our model, the inflationary observables $n_{\rm s}$, $n_{\rm t}$, $r$, and $\alpha_{\rm s}$ are expressed as functions of the parameters and constants $\zeta$, $\kappa$, $N$, and $\gamma$ as follows
\begin{align}
	\label{ns}&n_{\rm s}=1-2\epsilon_1-\epsilon_2=1+\frac{3\zeta^2}{4\kappa}\frac{3A(N)-1}{\big(1+A(N)\big)^2},\\
	\label{nt}&n_{\rm t}=-2\epsilon_{1}=\frac{3\zeta^2A(N)}{2\kappa\big(1+A(N)\big)^2},\\
	\label{r}&r=16\epsilon_1=-\frac{12\zeta^2A(N)}{\kappa\big(1+A(N)\big)^2},\\
	\label{alphas}
	&\alpha_{\rm s}=-2\epsilon_1\epsilon_2-\epsilon_2\epsilon_3=\frac{9\zeta^4}{16\kappa^2}\frac{4A(N)-3A^2(N)+1}{\big(1+A(N)\big)^4}.
\end{align}
In this context, the most recent constraints established by the Planck collaboration on the scalar spectral index, the tensor-to-scalar ratio, and the running of the scalar spectral index are presented in \cite{Planck:2018jri}, as follows
\begin{align}\label{planckdata}
	&n_{\rm s}= 0.9649\pm 0.0042\quad \nonumber\\
	&{}\qquad\left({\rm at}\ 68 \%\, {\rm CL, Planck TT,TE,EE+lowE+lensing}\right),\nonumber\\
	&\alpha_{\rm s}=- 0.0045\pm 0.0067\quad \nonumber\\
	&{}\qquad\left({\rm at}\ 68 \%\, {\rm CL, Planck TT,TE,EE+lowE+lensing}\right),\nonumber\\
	& r<0.10 \quad\left({\rm at}\ 95 \%\, {\rm CL, Planck
		TT+lowE+lensing}\right).
\end{align}

\begin{figure}[htbp!] 
	\centering 
	\includegraphics[width=3.5in]{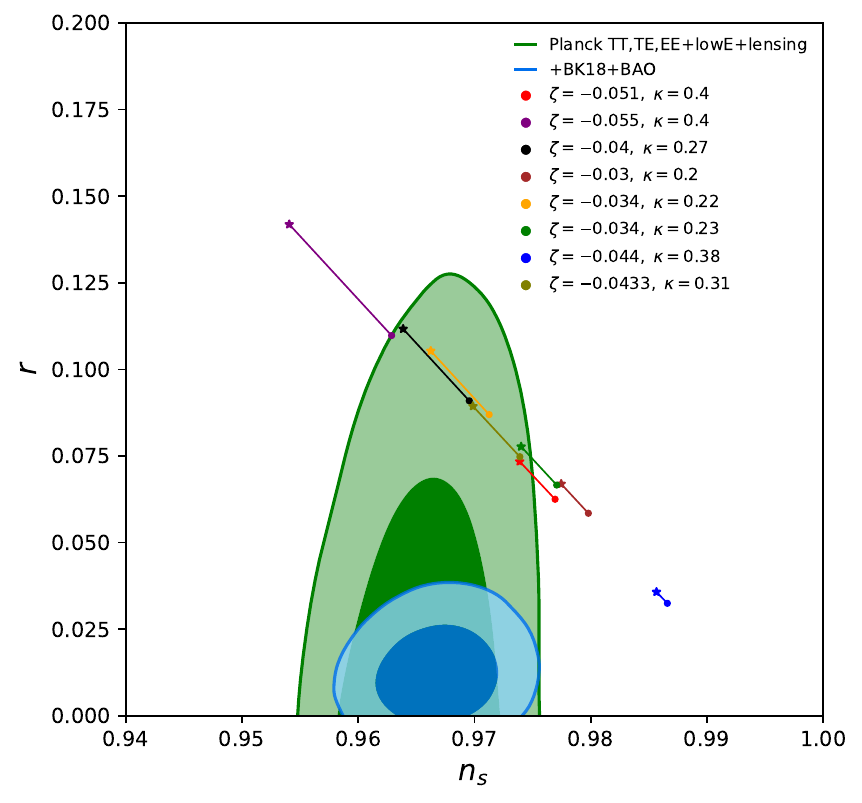}
	\caption{\label{fig4}Two-dimensional marginalized posterior distributions (68\% and 95\% confidence levels) in the $(n_s, r)$ plane are shown for different CMB datasets:  Planck TT,TE,EE+lowE+lensing (green), and Planck TT,TE,EE+lowE+lensing+BK18+BAO (blue). Overlaid are theoretical predictions from the considered inflationary model, parameterized by $(\zeta,\kappa)$. For each parameter pair, predictions at $N=50$ and $N=60$ e-folds are connected by lines, with circles marking $N=50$ and stars marking $N=60$. Different colors correspond to distinct $(\zeta,\kappa)$ values as indicated in the legend.	
	}
\end{figure}

\begin{figure}[htbp!] 
	\centering 
	\includegraphics[width=3.5in]{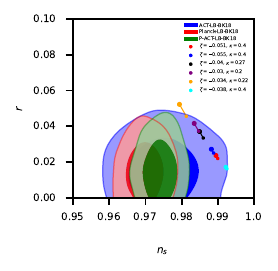}
	\caption{\label{fig16}Comparison of 2D  contours in the $(n_s, r)$ plane from ACT-LB-BK18 (blue), Planck-LB-BK18 (red), and combined P-ACT-LB-BK18 (green) datasets. Colored points indicate theoretical predictions for different parameter choices $(\zeta, \kappa)$ with fixed $\gamma=-1$ and $N=50$ (small points), $N=60$ (large points). Each pair of points is connected by a straight line to illustrate the evolution with e-folding number $N$. Red, blue, black, purple, orange, and cyan points correspond to $(\zeta, \kappa) = (-0.051, 0.4)$, $(-0.055, 0.4)$, $(-0.04, 0.27)$, $(-0.03, 0.2)$, $(-0.034, 0.22)$, and $(-0.038, 0.4)$, respectively. The plot highlights the relative positions of theoretical models with respect to observational constraints.}
\end{figure}

However, by conducting a joint analysis of the Planck, BK$18$, and BAO data \cite{BICEP:2021xfz}, further constraints have been imposed, which tighten the upper limit of $r$ to the specific value given by 
\begin{equation}\label{planckdata2}
	r<0.036\quad{\rm at}\ 95 \%\, {\rm CL}.
\end{equation}
Moreover, the recent constraints from the Atacama Cosmology Telescope (ACT) Data Release 6 \cite{ACT:2025fju,ACT:2025tim} and the joint analysis with BK18 \cite{BICEP:2021xfz} are given by
\begin{align}\label{planckdata}
	& n_{\rm s} = 0.9743 \pm 0.0034, \quad 
	\left({\rm at}\ 68\% \, {\rm CL, \ P\!-\!ACT\!-\!LB}\right), \notag\\
	& \alpha_{\rm s} = 0.0062 \pm 0.0052, \quad 
	\left({\rm at}\ 68\% \, {\rm CL, \ P\!-\!ACT\!-\!LB}\right), \notag\\
	& r < 0.038, \quad 
	\left({\rm at}\ 95\% \, {\rm CL, \ P\!-\!ACT\!-\!LB\!-\!BK18}\right).
\end{align}
\\
In Figs.~\ref{fighh}, \ref{figh}, \ref{figffh}, and \ref{figdd}, we present the scalar spectral index $n_{s}$, the tensor spectral index $n_{t}$, the tensor-to-scalar ratio $r$, and the running of the scalar spectral index $\alpha_{s}$ as functions of the model parameters $\zeta$ and  $\kappa$ for $N=70$ and $ \gamma=-1$. The results show that for certain values of $\zeta$ and $\kappa$, the model's predictions are in good agreement with the Planck 2018 data, the combined Planck$+$BK18$+$BAO datasets, as well as the P-ACT-LB and P-ACT-LB-BK18 measurements.
 Furthermore, in Fig.~\ref{fig4}, we plot the tensor-to-scalar ratio for  Planck$+$LowE$+$Lensing, and Planck$+$LowE$+$Lensing$+$BK18$+$BAO, as functions of the model parameters $\zeta$ and $ \kappa$ for $N=50,60$ and $\gamma=-1$.
These results indicate that for specific values of $\zeta$ and $\kappa$, the model's predictions are in close agreement with the Planck 2018 data as well as the combined Planck$+$BK18$+$BAO datasets.
\\
In Fig.~\ref{fig16}, we display the tensor-to-scalar ratio for ACT-LB-BK18, Planck-LB-BK18, and P-ACT-LB-BK18, as functions of the model parameters $\zeta$ and $\kappa$ for $N=50,60$ with $\gamma=-1$. 
Our results indicate that for specific values of $\zeta$ and $\kappa$, the model yields predictions that are in close agreement with the observational constraints from ACT-LB-BK18, Planck-LB-BK18, and P-ACT-LB-BK18.
\\
These findings indicate that the predictions of the Mori-Zwanzig formalism for the early universe are close to recent observational data, although they are not fully consistent with all measurements. Therefore, further modifications of the model are required to achieve full consistency with observational constraints.

\section{Potential Extensions and Open Questions}

\subsection{Potential of Scalar Field with Memory}
In this section, We aim to explore the possible existence of a memory-dependent scalar field along with its associated potential function. To show it paratactically, let's consider Einstein's field equations in the presence of a scalar field as follows
\begin{equation}
	G_{\mu\nu} =  T_{\mu\nu}^{(\phi)}
\end{equation}
where $G_{\mu\nu}$ is the Einstein tensor, and $T_{\mu\nu}^{(\phi)}$ is the energy-momentum tensor of the scalar field $\phi$. The energy-momentum tensor for a canonical scalar field is given by
\begin{equation}
	T_{\mu\nu}^{(\phi)} = \partial_\mu \phi \, \partial_\nu \phi - g_{\mu\nu} \left( \frac{1}{2} g^{\alpha\beta} \partial_\alpha \phi \, \partial_\beta \phi + V(\phi) \right).
\end{equation}
Additionally, the equation of motion for the scalar field follows from the Klein-Gordon equation in curved spacetime
\begin{equation}
	\Box \phi = \frac{dV(\phi)}{d\phi}.
\end{equation}
For the FLRW metric, the preceding equations take the form
\begin{align}
	\label{wwefdjwd}
&3H^{2}(t)=\frac{1}{2}\dot{\phi}^2(t)+V(\phi)\\
& \ddot{\phi}(t)+3H(t)\dot{\phi}(t)=-V_{,\phi}(\phi)
\end{align}
one can rewrite the above equations (see Ref.~\cite{Padmanabhan:2010zzb}) as follows
\begin{align}
	\label{fvefve}
V(t)=3H^2(t)+\dot{H}(t), \quad \text{with} \quad \phi(t)=\int dt \big(-2\dot{H}\big)^{1/2}.
\end{align}
Substituting the simplified form of  Eq.~\eqref{nnej} into Eq.~\eqref{fvefve} yields
\begin{align}
	\label{cvdfvadfv}
	&V(t)=\frac{3}{2}H^2(t)-\kappa \int_{0}^{t} dse^{-\zeta s}H(t-s), \\
	& \phi(t)=\int dt \Big(3H^2(t)+2\kappa \int_{0}^{t} dse^{-\zeta s}H(t-s)\Big)^{1/2}
	\label{dkjwdcjwe}
\end{align}
These describe a memory-dependent scalar field and its associated potential function, which is also memory-dependent. These equations should be studied more in different epochs of cosmology. For example, considering the scale factor $a(t) \propto t^n$, Eqs.~\eqref{cvdfvadfv} and \eqref{dkjwdcjwe} reduce to
\begin{align}
	\label{rghrghgr}
	&V(t)=\frac{3}{2}\frac{n^{2}}{t^2}-\kappa n e^{-\zeta t} \, \text{Ei}(\zeta t), \\
	& \phi(t)=\int dt \bigg(\frac{3n^2}{t^2}+2\kappa n e^{-\zeta t} \, \text{Ei}(\zeta t)\bigg)^{1/2}
	\label{fggrghergh}
\end{align}
where the exponential integral $\text{Ei}(x)$ is defined in Appendix \ref{wckjwcdwdn}. In the limit of long times, where $t \gg \frac{1}{\zeta}$, as indicated in Eq.~\eqref{wdckwfvf}, the exponential integral approximates to  $\text{Ei}(\zeta t) \simeq e^{\zeta t}/\zeta t$. Consequently, Eqs.~\eqref{rghrghgr} and \eqref{fggrghergh} can be expressed as
\begin{align}
	\label{dvvfdvfev}
	V(t)\simeq\frac{3}{2}\frac{n^{2}}{t^2}-\frac{\kappa n}{\zeta t}
\end{align}
with the scalar field given by
\begin{align}
\label{fvefefv}	
\phi(t)&\simeq \int dt \bigg(\frac{3n^2}{t^2}+\frac{2\kappa n}{\zeta t}\bigg)^{1/2}\notag\\
&\simeq 2\sqrt{3}n \left(\sqrt{1+\frac{2\kappa t}{3n\zeta}} - \operatorname{arccot}\left(\left(1+\frac{2\kappa t}{3n\zeta}\right)^{-1/2}\right)\right)
\end{align}	
 For long times $t \gg \zeta/\kappa$,  the scalar field function in Eq.~\eqref{fvefefv} simplifies further and ultimately becomes $\phi(t)\simeq 2\sqrt{2\kappa n t/\zeta}-\pi \sqrt{3}n$. Combining it with the potential function in Eq.~\eqref{dvvfdvfev}, the potential as a function of the scalar field takes the following form
 \begin{align}
 	\label{ffaevefv}
 	V(\phi)\simeq \frac{96 \kappa^2 n^4}{\zeta ^2}\big(\phi+\pi\sqrt{3}n\big)^{-4}-\frac{8\kappa^2 n^2}{\zeta ^2}\big(\phi+\pi\sqrt{3}n\big)^{-2}
 \end{align}
The potential functions of the form $V(\phi) \propto \phi^{-4}$ and $V(\phi) \propto \phi^{-2}$ have been derived from string theory in the context of D-brane inflation \cite{Kachru:2003sx, Dvali:1998pa}.\\
Moreover, one can consider another function for the scale factor, given by $a(t) \propto e^{\alpha t}$. In this case, the Hubble parameter is a constant, $H = \alpha$. Substituting this into Eqs.~\eqref{cvdfvadfv} and \eqref{dkjwdcjwe} yields
\begin{align}
	\label{efvcfeve}
&V(t)=\frac{3}{2}\alpha^2+\frac{\kappa \alpha}{\zeta}(e^{-\zeta t}-1)\\
& \phi(t)=\int dt \Big(3\alpha^2-\frac{2\kappa \alpha}{\zeta}(e^{-\zeta t}-1)\Big)^{1/2}
	\label{fverffr}	
\end{align}
For the limit $t \ll 1/\zeta$,  the scalar field described by Eq.~\eqref{fverffr} can be approximated as
$
\phi(t) \simeq (3\alpha^2 + 2\kappa \alpha t)^{3/2}/3\kappa \alpha.
$
Substituting this expression into Eq.~\eqref{efvcfeve} yields the potential function as a function of the scalar field, given by
\begin{equation}\label{wddcndnf}
V(\phi)\simeq 3\alpha^2-\frac{(3\kappa\alpha)^{2/3}}{2}\phi^{2/3}
\end{equation}
Power-law potentials of the form $V(\phi) \propto \phi^m$ have been widely explored in the study of single-field inflation. Commonly considered values of $m$, such as $m \in \{3, 2, 4/3, 1, 2/3\}$, have been tested against experimental data \cite{Planck:2018jri}.
 \\
In this section, we derived a power-law potential with $m = 2/3$ after applying certain simplifications and conditions. It has been shown that models with $m=1$ and $m=2/3$ are more compatible with the
data \cite{Silverstein:2008sg,McAllister:2008hb,Planck:2018jri}. 
Moreover, the potential function with $m = 2/3$ has been derived from string theory compactified on twisted tori \cite{Silverstein:2008sg,McAllister:2008hb} and is considered a candidate for describing the inflationary phase of the universe within string theory.\\  
If we could solve the exact form of Eq.~\eqref{efvcfeve}, the exact potential could be found as a function of the scalar field. However, the potential function in Eq.~\eqref{wddcndnf} is an important result derived in the short-time limit. 
The open question in this section is how to determine exact potential functions for scale factors of the forms $a(t) \propto e^{\alpha t}$ and $a(t) \propto t^n$. More generally, it involves finding various potential functions corresponding to different forms of the Hubble parameter by combining Eqs.~\eqref{cvdfvadfv} and \eqref{dkjwdcjwe}.
\\

One might inquire whether there are any known physical predictions where the scalar potential function takes an integral form, as shown in Eq.~\eqref{cvdfvadfv}. Indeed, such potential functions exist \cite{Weinberg:1973am}, including those found in grand unified theories (GUTs) \cite{Georgi:1974sy}. Here, I present one of these potential functions, known as the \textit{Coleman-Weinberg} potential, as given in \cite{Bardeen:1983qw,Albrecht:1982wi, Dolan:1973qd,Kirzhnits:1976ts, Linde:1978px,Billoire:1981fw}, and expressed as follows
\begin{align}
\label{dckjdcw}
V_{T}(\phi)=&(2\text{A}-\text{B})\sigma^{2}\phi^2-\text{A}\phi^4+\text{B}\phi^4\ln \Big(\frac{\phi^2}{\sigma^2}\Big)\notag\\
&+\frac{18T^4}{\pi^2}\int_{0}^{\infty}dx x^2\ln\Big(1-e^{-\sqrt{x^2+D(T)}}\Big)
\end{align}
where $T$ is the temperature,  $D(T)=25g^2\phi^2/(8 T^2)$, $\sigma = 4.5 \times 10^{14} \, \text{GeV}$, $\text{B}=5625 g^4/1024\pi^2$, and $\text{A}$ is a free parameter. The integral term on the right-hand side of Eq.~\eqref{dckjdcw} arises from the one-loop thermal correction of a bosonic excitation to the effective potential \cite{Dolan:1973qd}.\\
Therefore, the potential equation, formulated with an integral term, is well-established in fundamental physics and early cosmic inflation. However, the presence of a memory effect in fundamental interactions, as well as in the physics of cosmic inflation requires further investigation.

\subsection{Mori-Zwanzig  Stochastic Inflation}
The Mori-Zwanzig formalism provides a systematic approach to deriving the generalized Langevin equation (GLE) by projecting out fast degrees of freedom. It is discussed in  Section \ref{csdjcadc}.
\\
If the memory kernel $K(t)$ decays rapidly, we approximate it as a delta function
\begin{equation}
	K(t) \approx \gamma \delta(t).
\end{equation}
This simplifies Eq.~\eqref{sddbedbejw} to the standard Langevin equation
\begin{equation}\label{dcdcjdc}
	\frac{d A}{d t} = -\gamma A + F(t).
\end{equation}
For a velocity variable $v(t)$, this becomes
\begin{equation}
	m \frac{d v}{d t} = -\gamma v + \eta(t),
\end{equation}
where $\eta(t)$ is a Gaussian white noise with
\begin{equation}
	\langle \eta(t) \rangle = 0, \quad \langle \eta(t) \eta(t') \rangle = 2D \delta(t - t').
\end{equation}
The Mori-Zwanzig equation provides a non-Markovian description of system dynamics. Under the Markovian approximation, we recover the Langevin equation with a friction term and a stochastic force, consistent with the fluctuation-dissipation theorem. \\
This formalism provides a systematic way to separate the relevant (long-wavelength) dynamics from the irrelevant (short-wavelength) fluctuations. 	We decompose the field into long-wavelength and short-wavelength components
\begin{equation}\label{dkjwedjfwd}
	\phi(x,t) = \phi_l(x,t) + \phi_s(x,t).
\end{equation}
where,
\begin{itemize}
	\item 
	$\phi_{l}(x,t)$ contains long-wavelength modes ($k<k_{c}$), describing coarse-grained dynamics.
	\item $\phi_{s}(x,t)$ contains short-wavelength modes ($k>k_{c}$), which will be systematically integrated out.
\end{itemize}
In Fourier space the scaler field Eq.~\eqref{dkjwedjfwd} is written
\begin{equation}
\tilde{\phi}(k,t)=\Theta(k_{c}-k)\tilde{\phi}_{l}(k,t)+\Theta(k-k_{c})\tilde{\phi}_{s}(k,t)
\end{equation}
where $\Theta(k)$ is the Heaviside step function that select the appropriate modes.
\\
We define the projection operate $\mathcal{P}$ that extracts the long-wavelength dynamics as follows
\begin{equation}
\mathcal{P}f=\int_{|k|<k_{c}}dk\tilde{f}(k,t)e^{ikx}
\end{equation}
The complementary operator $\mathcal{Q}=1-\mathcal{P}$ extracts the short-wavelength components. Now the equation of motion can be written as 
\begin{equation}\label{wdfjwdfcje}
\frac{d}{dt}\phi=i\mathcal{L}\phi
\end{equation}	
where $i\mathcal{L}$ is the Liouville operator describing the field's evolution. We decompose Eq.~\eqref{wdfjwdfcje} into long and short modes
\begin{align}\label{dgbrghgrgh}
	\frac{d}{dt}\phi=i\mathcal{L}\mathcal{P}\phi+i\mathcal{L}\mathcal{Q}\phi
\end{align}	
This equation can be formally solved as (see Appendix \ref{dccldecnj})
\begin{equation}
\phi(t)=e^{i\mathcal{L}\mathcal{Q}(t-t_{i})}\phi(t_{i})+\int_{t_{i}}^{t}e^{i\mathcal{L}\mathcal{Q}(t-s)}i\mathcal{L}\mathcal{P}\phi(s)ds	
\end{equation}
we write this equation in the following form
\begin{equation}\label{cnwdcn}
\phi(t)=
\int_{t_{i}}^{t}K(t-s)\phi_{l}(s)ds+f(t)
\end{equation} 	
where $\phi_{l}(s)=\mathcal{P}\phi(s)$,  $K(t-s)=e^{i\mathcal{L}\mathcal{Q}(t-s)}i\mathcal{L}$ is the memory kernel, encoding the influence of short-scale modes on long-scale dynamics, and $f(t)=e^{i\mathcal{L}\mathcal{Q}(t-t_{i})}\phi(t_{i})$ is the noise term, representing unresolved degrees of freedom. 
\\
Now, taking the limit $K(t-s)\approx \delta(t-s)$,  we rewrite Eq.~\eqref{cnwdcn} as
\begin{equation}
	\phi(t)=
	\phi_{l}(t)+f(t)
\end{equation} 
This equation is similar to the one studied in the context of stochastic inflation \cite{Starobinsky:1986fx, Starobinsky:1994bd}, where $f(t)$ is defined as
\begin{equation}
f(t)=\frac{1}{(2\pi)^{3/2}}\int \Theta(k-k_{c})\big(a_{k}\phi_{k}(t)e^{ik.x}+a^{\dagger}_{k}\phi^{*}_{k}(t)e^{-ik.x}\big).
\end{equation} 
The open question in this section is whether it is possible to formulate a general equation to describe the dynamics of stochastic inflation, as an alternative to the Langevin equation, in the following form
\begin{equation}\label{dcwdfcdw}
	\frac{d}{dt}\phi_{l}(t)=\Omega\phi_{l}(t)+
	\int_{t_{i}}^{t}K(t-s)\phi_{l}(s)ds+F(t)
\end{equation}
For the approximation $K(t-s) \approx \gamma \delta(t-s)$, this equation reduces to the standard Langevin equation \eqref{dcdcjdc}, which has been extensively studied in the context of stochastic inflation \cite{Starobinsky:1986fx, Starobinsky:1994bd}. 
The presence of a non-local memory term in Eq.~\eqref{dcwdfcdw} could represent an enigmatic characteristic of our universe during the early inflationary phase. Further investigation is necessary to explore the possible existence of this memory property in the early universe.\\
Another perspective on the dependence of the time derivative of the scalar field on the memory term is provided in Eq.~\eqref{dkjwdcjwe}, where $\dot{\phi}\propto \int e^{-\zeta s}H(t-s)ds$. \\
This suggests that the time evolution of the scalar field, and more generally, the early universe, depends on the system's time history. I define this class of universes as the \textit{Cognitive Universe}, characterized by an evolution that is inherently dependent on its past states. In such a universe, the dynamics at any given moment are not solely determined by local conditions but are influenced by the accumulated history of the system. This memory-driven behavior, encoded through integral or non-local terms in the governing equations, distinguishes the Cognitive Universe from conventional cosmological models, where evolution is typically Markovian and dictated only by instantaneous parameters.
\section{Conclusions}\label{sec:iv}
In this paper, we introduced a memory-dependent equation of state (MDES) and explored its implications for cosmology, particularly in the early universe. By incorporating memory effects into the pressure function, we demonstrated that the equation of state naturally leads to an inflationary phase at high energy densities. Our formulation, based on integral memory terms, provides a more generalized description of thermodynamic evolution, distinguishing between the oblivion and memory contributions to pressure.
\\
To establish a deeper connection between nonequilibrium thermodynamics and cosmology, we applied the Mori-Zwanzig formalism to derive an effective evolution equation for the Hubble parameter. This approach led to a modified version of the Friedmann equations, incorporating memory terms that influence the expansion history of the universe. The resulting inflationary dynamics were analyzed through slow-roll parameters, confirming that our model is consistent with an early accelerated expansion. 
Furthermore, we compared our theoretical predictions with observational constraints from Planck, BAO, ACT, and BK18 data, demonstrating that our model provides a viable description of cosmic inflation. The spectral index $n_s$, tensor-to-scalar ratio $r$, and running of the spectral index $\alpha_s$ were computed within our framework, exhibiting close consistency with current observational constraints.
\\\\
These findings highlight the potential of memory-dependent equations of state in addressing fundamental problems in cosmology. Future work could extend this approach to late-time cosmology, investigating whether memory effects play a role in dark energy evolution and the dynamics of structure formation. Additionally, exploring the interplay between memory-dependent thermodynamics and modified gravity theories may provide further insights into the fundamental nature of spacetime and cosmological evolution.\\

\appendix
\section{Bernstein–Kearsley–Zapas model}
\label{dekwdwkj}
Let each particle in a solid material be labeled by a triplet $X^a$ (with $a = 1, 2, 3$), and let its spatial position be described by the coordinates $x^i$ (with $i = 1, 2, 3$) \cite{bernstein1963stress}. Therefore, the motion of a particle is described by
\begin{equation}\label{knwdcwkjdc}
x^i = x^i(X^a, t)
\end{equation}
The deformation gradient is defined as
\begin{equation}
	F_a^i = \frac{\partial x^i}{\partial X^a}
\end{equation}
Equation \eqref{knwdcwkjdc} asserts that each particle can occupy only one position in space at any given time. Additionally, we assume that no more than one particle exists at any spatial point at any moment. Consequently, Equation \eqref{knwdcwkjdc}  can be inverted as
\begin{equation}\label{sfvvcd}
	X^a =X^a (x^i, t)
\end{equation}
\\
The inverse deformation gradient is given by
\begin{equation}
	G^a_i = \frac{\partial X^a}{\partial x^i}.
\end{equation}
Assuming the material is incompressible, meaning that the volume of a system or a material element does not change during deformation or flow, we have
\begin{equation}\label{kjsdqskjbd}
\text{det}|F_a^i| = 1
\end{equation}
The right Cauchy-Green tensor $C_{ij}$ is defined as
\begin{equation}
C_{\alpha\beta} = F^i_\alpha F^j_\beta \delta_{ij}
\end{equation}
According to Equation \eqref{kjsdqskjbd}, it follows that $\text{det}|C_{\alpha\beta}| = 1$. The Green-St. Venant tensor 
$E_{\alpha\beta}$ is expressed as
\begin{equation}\label{wdjfwdfwei}
E_{\alpha\beta} = \frac{1}{2}(C_{\alpha\beta} - \delta_{\alpha\beta}).
\end{equation}
It is assumed that at any time the stress at a particle is determined, up to a hydrostatic pressure, by the values of the deformation gradients at that particle for all prior times \cite{bernstein1963study}. Indeed, taking into account the principle of material indifference, it is posited that there exists a functional of $x^{i}_{a}(s)$, where $-\infty<s<t$, which is denoted by $\Psi^{\alpha\beta}[E_{\mu\nu}(s), t]$.  
Therefore, 
The true stress tensor 
$\sigma_{ij}$ at any time $t$ can be represented as
\begin{equation}\label{dwcwdkjcd}
	\sigma^{ij}(t) = -p \delta^{ij} + F^i_{\alpha}(t) F^j_\beta(t) \Psi^{\alpha\beta}[E_{\mu\nu}(s), t]
\end{equation}
where $p$ is the hydrostatic pressure, $\delta_{ij}$ is the Kronecker delta, and $\Psi$ is a functional that depends on the material's deformation.\\
According to Green and Rivlin's theory \cite{Green1957,Green1959part2,Green1959}, the function $\Psi^{\alpha\beta}$ can be expressed as
\begin{align}\label{cdnwdlfwd}
\Psi^{\alpha\beta}&[E_{\mu\nu}(s), t]=\sum_{i=0}^{n}k^{\alpha\beta\mu_{1}\nu_{1}\dots \mu_{i}\nu_{i}}E_{\mu_{1}\nu_{1}}(t)\dots E_{\mu_{i}\nu_{i}}(t)
\notag\\
&-\sum_{i=1}^{n} \int_{-\infty}^{t} \dots  \int_{-\infty}^{t}K^{\alpha\beta\mu_{1}\nu_{1}\dots \mu_{i}\nu_{i}}(t-s_{1},\dots, t-s_{i})\notag\\
&\quad\quad\quad\quad\quad\quad \quad\quad \times E_{\mu_{1}\nu_{1}}(s)\dots E_{\mu_{i}\nu_{i}}(s)ds_{1}\dots ds_{i}
\end{align}
Substituting the memory-dependent function into Eq.~\eqref{dwcwdkjcd} reveals that the present value of the stress tensor is influenced by the system's evolutionary history
\section{Perfect elastic fluid}
\label{dkldefkddfni}
In the study of inviscid (non-viscous) gases, the caloric equation of state defines the specific internal energy $u$ as a function of the specific volume $v$ and specific entropy $h$ as: $u=f(v,h)$. The functions that relate local pressure $p_{eq}$ and local absolute temperature $T$ to specific volume and specific entropy are expressed as: $p_{eq}=-f_{v}(v, h)=-\partial f/\partial v$ and $T=f_{h}(v, h)=\partial f/\partial h$, where one can write this relation $du=Tdh-p_{eq}dv$. This equations are define in \textit{local thermodynamic equilibrium}, in the sense that, a given volume of gas may not always be in equilibrium, as properties like temperature and pressure can vary across different points. However, at each point, local thermodynamic equilibrium is maintained.\\
When viscoelastic behavior is considered \cite{bernstein1964thermodynamics}, it is viewed that such behavior cannot be explained by local thermodynamic equilibrium, and that shear stress cannot be supported by a fluid in thermodynamic equilibrium. Therefore, people have tried to create a theory to explain fluids that are not in local thermodynamic equilibrium.\\
In this case, the caloric equation of state is defined as
\begin{equation}
u=f(v,h+\Sigma)
\end{equation}
where a non-negative quantity $\Sigma$, with the dimensions of entropy per unit mass, that vanishes at thermodynamic equilibrium. In fact, local equilibrium at a point occurs if and only if $\Sigma= 0$ at that point. The temperature is expressed as $T=f_{h}(v, h+\Sigma)$, while the pressure is given by
\begin{equation}\label{dcwddcwdidcw}
p=-f_{v}(v,h+\Sigma)
\end{equation}
It has been demonstrated \cite{bernstein1964thermodynamics} that the present-time value of $\Sigma$ depends on the past history of deformation and temperature, as given by
\begin{equation}
\Sigma(t)=\int^{t}_{-\infty}S\big[\boldsymbol{E}(t,\tau),B(t, \tau)\big]b_{T}(\tau)d\tau	
\end{equation}
where $\boldsymbol{E}(t,\tau)$ is the strain tensor defined in Eq.~\eqref{wdjfwdfwei}, and $B(t,\tau)=\int^{t}_{\tau}b_{T}(\xi)d\xi$. The quantity $b_{T}$ depends on the absolute temperature $T$ and equals unity at the reference temperature $T_{0}$. \\
Given the previous discussions, the stress tensor is defined as

\begin{align}\label{dfgnbgfbfg}
\sigma^{ij}(t)=	-p \delta^{ij}
	+\rho T \int_{-\infty}^{t}&ds~F^{i}_{\alpha}(t,s)F^{j}_{\beta}(t,s)b_{T}(s)
	\notag\\
	&\times S^{\alpha\beta}\Big[\boldsymbol{E}(t,s),\int^{t}_{s}b_{T}(\eta)d\eta\Big]
\end{align} 
where $S^{\alpha\beta}=\partial S/\partial E_{\alpha\beta}$.

\section{Entropy Balance and the Effective Work Term}\label{appendixcc}

We adopt the following sign conventions: $\delta Q$ is the heat supplied to the system, $\delta W$ is the work done \emph{by} the system, and $dU$ is the change of internal energy. The first law of thermodynamics reads
\begin{equation}\label{eq:firstlaw}
	dU = \delta Q - \delta W.
\end{equation}

The entropy balance (local Clausius relation including irreversible production) is
\begin{equation}\label{eq:entbal}
	dS = \frac{\delta Q}{T} + d_i S, 
	\qquad d_i S \ge 0,
\end{equation}
where $d_i S$ denotes internal (irreversible) entropy production, e.g., due to viscous dissipation.

\paragraph{Reversible case.} 
For a simple compressible system in local thermodynamic equilibrium, the Gibbs (fundamental) relation gives
\begin{equation}\label{eq:gibbs}
	T\,dS = dU + p_{e}\,dV,
\end{equation}
where $p_{e}$ is the equilibrium pressure and $V$ the volume. For a reversible process, $d_i S = 0$. Combining \eqref{eq:firstlaw}, \eqref{eq:entbal} (with $d_iS=0$), and \eqref{eq:gibbs} gives
\begin{equation}\label{eq:reversible_entropy}
	T\,dS = \delta Q \quad \Rightarrow \quad dS = \frac{\delta Q}{T},
\end{equation}
and
\begin{equation}\label{eq:reversible_work}
	\delta W = \delta Q - dU = T\,dS - dU = p_{e}\,dV.
\end{equation}

\paragraph{Irreversible (viscous) case.} 
When irreversible viscous processes are present, the entropy balance \eqref{eq:entbal} has $d_i S \neq 0$. Substituting $\delta Q$ from \eqref{eq:firstlaw} into \eqref{eq:entbal} yields
\begin{equation}\label{eq:TdS_irrev}
	dS = \frac{dU + \delta W}{T} + d_i S
	\quad \Longrightarrow \quad
	T\,dS = dU + \delta W + T\,d_i S.
\end{equation}

Comparing this expression for $T\,dS$ with the Gibbs relation \eqref{eq:gibbs} for local equilibrium variables,
\begin{equation}\label{eq:gibbs_eq}
	T\,dS = dU + p_{e}\,dV,
\end{equation}
we equate the two forms of $T\,dS$:
\begin{equation}\label{eq:compare_TdS}
	dU + \delta W + T\,d_i S = dU + p_{e}\,dV.
\end{equation}
Canceling $dU$ on both sides and rearranging gives
\begin{equation}\label{eq:deltaW}
	\delta W = p_{e}\,dV - T\,d_i S.
\end{equation}
We define the \emph{bulk viscous pressure} $\Pi$ through
\begin{equation}\label{eq:Pi_def}
	T\,d_i S \equiv -\,\Pi\,dV,
\end{equation}
so that the work contribution can be written as
\begin{equation}\label{eq:deltaW_bulk}
	\delta W = (p_{e} + \Pi)\,dV \equiv p^{\text{eff}}\,dV,
\end{equation}
with the effective pressure $p^{\text{eff}}$. Explicitly,
\begin{equation}\label{eq:Pi_explicit}
	\Pi = -\,T\,\frac{d_i S}{dV},
\end{equation}
where $d_i S/dV$ is the irreversible entropy production per unit volume. The sign convention ensures $d_i S \ge 0$, so that for an expanding fluid $\Pi$ is typically negative, reducing the effective mechanical pressure.

\section{Equation of motion from Mori-Zwanzig formalism}\label{sdqdnwd}
According to the Heisenberg equation of motion, $\frac{dA}{dt}=\frac{i}{\hbar}[H,A]=iLA$, we can express the time evolution of the operator $A$ as $A(t)=e^{iLt}A$. From this, the following definitions can be established
\begin{equation}\label{djeejbd23}
H(t)=e^{iLt}H,~~\dot{H}(t)=e^{iLt}\dot{H},~~\ddot{H}(t)=e^{iLt}\ddot{H}
\end{equation}	
By applying the operator $e^{iLt}$ to the microscopic equation of motion Eq.~\eqref{djediede}, and utilizing equation Eq.~\eqref{djeejbd23} along with the Dyson identity Eq.~ \eqref{sdbjebdei}, we arrive at the following equation
\begin{equation}\label{dnwjedwe}
\dot{H}(t)=-\frac{3}{2}H^{2}(t)+e^{iLt}Q\Psi+e^{iLt}P\Psi
\end{equation}	
where $P+Q=1$, and the above terms are defined as 
\begin{equation}\label{dnwewe}
e^{iLt}P\Psi=A_{j}(t)	(A_{j},A_{k})^{-1}(\Psi,A_{k})=A_{j}(t)\Omega_{A_{j}H}
\end{equation}
\begin{align}\label{dcbWDBISWI}
	e^{iLt}Q\Psi&=F_{H}(t)+\int^{t}_{0}dse^{iL(t-s)}PiLF_{H}(t)\notag\\
	&=F_{H}(t)+\int^{t}_{0}dsA_{j}(t-s)K_{ij}(s)
\end{align}	
with $F_{H}(t)=e^{iQLt}Q\Psi$,   $\Omega_{A_{j}H}=(A_{j},A_{k})^{-1}(\Psi,A_{k})$, and $K_{ij}(s)=(A_{j},A_{k})^{-1}(iLF_{H}(t),A_{k})$.\\
Substituting Eqs.~ \eqref{dcbWDBISWI} and \eqref{dnwewe} into Eq.~ \eqref{dnwjedwe} results in the following equation
\begin{align}\label{dbwedjbwe}
\dot{H}(t)=&-\frac{3}{2}H^{2}(t)+\Omega_{HH}H(t)+\Omega_{HH^2}H^{2}(t)\notag\\&+\int^{t}_{0}ds\big[K_{HH}(s)H(t-s)+K_{HH^2}(s)H^{2}(t-s)\big]\notag\\
&+F_{H}(t)
\end{align}	
where $F_{H}(t)$ is the noise term. It is assumed that noise term is small and can be omitted. 
Considering some simplifications $\Omega_{HH}=\Omega_{HH^2}=0$, $K_{HH^2}(s)=0$, and $K_{HH}(s)=-\kappa e^{-\zeta s}$ one can write Eq.~\eqref{dbwedjbwe} in the following form
 \begin{align}\label{sdbwdwe}
 	\dot{H}(t)=-\frac{3}{2}H^{2}(t)-\kappa\int^{t}_{0}ds e^{-\zeta s}H(t-s)
 \end{align}	
taking the time derivative of the above equation yields a new differential equation
\begin{align}\label{sdndqwwd} 
\ddot{H}(t)=&-3\dot{H}(t)H(t)-\kappa e^{-\zeta t}H(0)\notag\\
&+\kappa\int^{t}_{0}ds e^{-\zeta s}\frac{\partial H(t-s)}{\partial s}
\end{align}
Finally, using the  integration by parts Eq.~ \eqref{sdndqwwd} is written 
\begin{equation}\label{sdjdnwqn}
\ddot{H}(t)=-3\dot{H}(t)H(t)-\xi\Big(\dot{H}(t)+\frac{3}{2}H^2(t)\Big)-\kappa H(t).
\end{equation}
This is the main equation we have worked on to study the early cosmic inflation. It is a third-order differential equation with respect to the effective scale factor, and finding a general solution is not straightforward.
\section{Slow-roll parameters}
\label{dsbdddjowbd}
Consider the standard freedman equations  
\begin{subequations}
\begin{align}\label{fgbrghbgfb}
&H^2+\frac{k}{a^2}=\frac{8\pi G}{3}\rho\\
&\dot{H}+H^2=-\frac{4\pi G}{3}(\rho+3p)
\label{ddndnkpskw}
\end{align}
\end{subequations}
for a spatially flat spacetime with $k = 0$ and an equation of state given by $ p = \omega \rho$, the above equations can be expressed as
\begin{subequations}
\begin{align}\label{dedvdv}
&H^2=\frac{8\pi G}{3}\rho\\
&H^2\Big(1+\frac{\dot{H}}{H^2}\Big)=-\frac{4\pi G}{3}(1+3\omega)\rho
\label{fvcevfed}
\end{align}
\end{subequations}
combining these two equations leads to the following conditions
\begin{equation}\label{dfkwddfbwqe}
	\text{For} \quad \frac{\dot{H}}{H^2}\ll 1, \quad  \text{we have} \quad \omega\simeq -1.
\end{equation}

Taking the time derivatives of Eqs.~\eqref{dedvdv} and \eqref{fvcevfed}  leads to
\begin{subequations}
\begin{align}\label{fvfevfe}
&2H\dot{H}=\frac{8\pi G}{3}\dot{\rho}\\
&2H\dot{H}\Big(1+\frac{\ddot{H}}{2H\dot{H}}\Big)=-\frac{4\pi G}{3}(1+3\omega)\dot{\rho}
\label{fevefvv}
\end{align}
\end{subequations}
combining these two equations results in the following conditions
\begin{equation}\label{gbfgfdc}
	\text{For} \quad \frac{\ddot{H}}{2H\dot{H}}\ll 1, \quad  \text{we have} \quad \omega\simeq -1.
\end{equation}
The conditions \eqref{dfkwddfbwqe} and \eqref{gbfgfdc} are known as slow-roll parameters. In general the $(n + 1)$st Hubble slow-roll parameter $(n \geq 1)$ can be defined as
\begin{equation}\label{djndcjw}
\epsilon_{n+1}=	\frac{\dot{\epsilon}_{n}}{H\epsilon_{n}}, \quad \quad \text{with} \quad \epsilon_1\equiv -\frac{\dot{H}}{H^2}. 
\end{equation}	
\\\
The slow-roll parameters as function of Hubble parameter and its time derivatives are given by
\begin{align}
\label{eps1H}&\epsilon_1\equiv -\frac{\dot{H}}{H^2},\\
\label{eps2H}&\epsilon_2\equiv \Bigg\lvert\frac{\ddot{H}}{H\dot{H}}-2\frac{\dot{H}}{H^2}\Bigg\rvert,\\
&\epsilon_3\equiv \frac{1}{\left(\ddot{H}H-2\dot{H}^2\right)}\nonumber\\
\label{eps3H}&\hspace*{0.4cm}\times \Bigg(\dddot{H}-\frac{\ddot{H}\dot{H}}{H}-\frac{\ddot{H}^2}{\dot{H}}-\frac{2\dot{H}\ddot{H}}{H}+\frac{4\dot{H}^3}{H^2}\Bigg)
\end{align}
The slow-roll parameters \eqref{eps1H}, \eqref{eps2H}, and \eqref{eps3H} can be reexpressed as functions of the e-folding number as follows
\begin{align}
	\label{eps1N}&\epsilon_1(N)=-\frac{H^\prime(N)}{H(N)},\\
	\label{eps2N}&\epsilon_2(N)=\Bigg\lvert\frac{H^{\prime\prime}(N)}{H^\prime(N)}-\frac{H^\prime(N)}{H(N)}\Bigg\rvert\\
	&\epsilon_3(N)=\bigg(\frac{H(N)H^\prime(N)}{H(N)H^{\prime\prime}(N)-H^{\prime 2}(N)}\bigg)\nonumber\\
	\label{eps3N}
	&\hspace*{1cm}\times \bigg(\frac{H^{\prime\prime\prime}(N)}{H^{\prime}(N)}-\frac{H^{\prime\prime 2}(N)}{H^{\prime 2}(N)}-\frac{H^{\prime\prime}(N)}{H(N)}+\frac{H^{\prime 2}(N)}{H^2(N)}\bigg).
\end{align}

\section{CPL-like Parametrization of $\omega(\rho)$ Centered at $a_i$}
\label{dmdwdcw}

In this appendix, we derive a linear parametrization of the dark-energy equation-of-state
parameter $\omega$ as a function of the scale factor $a$, starting from the
density-dependent form
\begin{equation}
	\omega(\rho)\simeq -1 + \frac{\zeta}{\sqrt{3\rho}} + \frac{2\kappa}{3\rho},
\end{equation}
where $\zeta$ and $\kappa$ are model parameters.\\

We consider a generalized Chevallier-Polarski-Linder (CPL) expansion centered at an arbitrary
scale factor $a_i$
\begin{equation}
	w(a) = w_0 + w_a (a_i - a),
\end{equation}
where $w_0$ is the value of $\omega$ at $a = a_i$
\begin{equation}
	w_0 \equiv \omega(\rho_i), \qquad \rho_i \equiv \rho(a_i),
\end{equation}
and $w_a$ encodes the slope
\begin{equation}
	w_a \equiv - \frac{d\omega}{da}\Big|_{a=a_i}.
\end{equation}
Using the chain rule,
\begin{equation}
	\frac{d\omega}{da} = \frac{d\omega}{d\rho}\frac{d\rho}{da},
\end{equation}
and assuming a self-conserved dark-energy component, the continuity equation gives
\begin{equation}
	\frac{d\rho}{da} = -\frac{3(1+\omega(\rho))\rho}{a}.
\end{equation}
Evaluated at $a = a_i$
\begin{equation}
	\frac{d\rho}{da}\Big|_{a=a_i} = -\frac{3\,\delta_i\,\rho_i}{a_i}, \qquad \delta_i \equiv 1 + \omega(\rho_i).
\end{equation}
The derivative of $\omega$ with respect to $\rho$ is
\begin{equation}
	\frac{d\omega}{d\rho} = -\frac{\zeta}{2\sqrt{3}} \rho^{-3/2} - \frac{2\kappa}{3} \rho^{-2},
\end{equation}
and evaluated at $\rho_i$
\begin{equation}
	\frac{d\omega}{d\rho}\Big|_{\rho_i} = -\frac{\zeta}{2\sqrt{3}\,\rho_i^{3/2}} - \frac{2\kappa}{3\,\rho_i^{2}}.
\end{equation}
Combining the above results, the slope parameter is
\begin{equation}
	\begin{aligned}
		w_a &= -\frac{d\omega}{da}\Big|_{a=a_i} 
		= -\frac{d\omega}{d\rho}\Big|_{\rho_i} \frac{d\rho}{da}\Big|_{a=a_i} \\
		&=- \frac{3\,\delta_i\,\rho_i}{a_i} \left( \frac{\zeta}{2\sqrt{3}\,\rho_i^{3/2}} + \frac{2\kappa}{3\,\rho_i^{2}} \right)
	\end{aligned}
\end{equation}
with
\begin{equation}
	\delta_i =  1 + \omega(\rho_i) = \frac{\zeta}{\sqrt{3\rho_i}} + \frac{2\kappa}{3\rho_i}.
\end{equation}
Thus, the generalized CPL-like equation-of-state parametrization centered at $a_i$ reads
\begin{equation}
	\begin{aligned}
		w(a) &= w_0 + w_a (a_i - a), \\
		w_0 &= -1 + \frac{\zeta}{\sqrt{3\rho_i}} + \frac{2\kappa}{3\rho_i}, \\
		w_a &= -\frac{3\,\delta_i\,\rho_i}{a_i} 
		\left( \frac{\zeta}{2\sqrt{3}\,\rho_i^{3/2}} + \frac{2\kappa}{3\,\rho_i^{2}} \right)
	\end{aligned}
\end{equation}
This form can be used to model a dark-energy component whose equation of state evolves linearly with the scale factor around an arbitrary reference point $a_i$.

\section{Calculations in the Mori-Zwanzig Formalism}
\label{dccldecnj}
The equation for $\phi$ is
\begin{equation}\label{dfwefweq}
\frac{d}{dt}\phi=i\mathcal{L}\mathcal{P}\phi+i\mathcal{L}\mathcal{Q}\phi
\end{equation}	
this is the first-order linear differential equation, which we solve using the integrating factor method. Defining the integrating factor $U(t)=e^{-i\mathcal{L}\mathcal{Q}t}$, and multiplying both sides of Eq.~\eqref{dfwefweq} by $U(t)$, we get
\begin{equation}\label{hhcchctu}
	U(t)\frac{d}{dt}\phi=U(t)i\mathcal{L}\mathcal{P}\phi+U(t)i\mathcal{L}\mathcal{Q}\phi
\end{equation}
the above equation can be rewritten as
\begin{equation}
\frac{d}{dt}\big(U(t)\phi\big)=U(t)i\mathcal{L}\mathcal{P}\phi
\end{equation}
integrating both sides of the equation above from $t_{i}$ to $t$ yields
\begin{equation}
U(t)\phi(t)-U(t_{i})\phi(t_{i})=\int_{t_{i}}^{t}U(s)i\mathcal{L}\mathcal{P}\phi(s)ds
\end{equation}
multiplying both sides by 
$e^{i\mathcal{L}\mathcal{Q}t}$ results in
\begin{equation}
\phi(t)=e^{i\mathcal{L}\mathcal{Q}(t-t_{i})}\phi(t_{i})+\int_{t_{i}}^{t}e^{i\mathcal{L}\mathcal{Q}(t-s)}i\mathcal{L}\mathcal{P}\phi(s)ds
\end{equation}
\section{Lambert Function}
\label{rtergttr}
The Lambert W function \cite{Corless1996}, denoted as $W(x)$, is defined by the equation
\begin{equation}
x = W(x) e^{W(x)}.
\end{equation}
The function has multiple branches, with the principal branch denoted as $W_0(x)$ and the secondary branch denoted as  $W_{-1}(x)$. Some important properties include
\begin{itemize}
\item For $x \geq 0$, 
$ W_0(x)
$ is the real-valued function.
\item For $-\frac{1}{e} \leq x < 0
$, $W_{-1}(x)$ is real.
\item Special values include
\begin{equation}
W(0)  = 0, \quad \text{and} \quad
W\left(-\frac{1}{e}\right) = -1
\end{equation}
\end{itemize}
Given the Lambert $W$ function applied to a general function $f(x)$, i.e., $W(f(x))$, the derivative of $W(f(x))$ can be derived. 
Start with the derivative of the Lambert $W$ function
\begin{equation}
\frac{dW(u)}{du} = \frac{W(u)}{u(1 + W(u))},
\end{equation}
where $u = f(x)$.
By using the chain rule
\begin{equation}
\frac{d}{dx} W(f(x)) = \frac{dW(f(x))}{df(x)} \cdot \frac{df(x)}{dx}.
\end{equation}
and substitute the derivative of $W(u)$, the final result is given as 
\begin{equation}
\frac{d}{dx} W(f(x)) = \frac{W(f(x))}{(1 + W(f(x)))} \frac{f'(x)}{f(x)}.
\end{equation}
\\

\section{Exponential Integral}
\label{wckjwcdwdn}
The exponential integral \cite{Arfken2013}, denoted as $\text{Ei}(x)$, is defined for $x < 0$ as
\begin{equation}
	\text{Ei}(x) = -\int_{-x}^{\infty} \frac{e^{-u}}{u} \, du.
\end{equation}
For $x > 0$, the exponential integral is defined as
\begin{equation}
	\text{Ei}(x) = \int_{-\infty}^{x} \frac{e^{u}}{u} \, du.
\end{equation}
The exponential integral has several important properties. 
For small values of $x$, $\text{Ei}(x)$ can be approximated by the series
\begin{equation}
	\text{Ei}(x) = \gamma + \ln(x) + \sum_{n=1}^{\infty} \frac{x^n}{n \cdot n!}.
\end{equation}
where $\gamma$ is the Euler-Mascheroni constant.
For large positive $x$
\begin{equation}\label{wdckwfvf}
	\text{Ei}(x) \sim \frac{e^x}{x} \quad \text{as } x \to \infty.
\end{equation}
The exponential integral is widely used in various fields, including physics and engineering, particularly in problems involving exponential decay and integrals involving logarithmic singularities.

\bibliography{apssamp}
\end{document}